\documentclass[12pt]{article}
\usepackage{amsbsy}
\usepackage{amsfonts}
\usepackage{amsmath}
\usepackage{amssymb}
\usepackage{apacite}
\usepackage{setspace}
\usepackage{graphicx}
\usepackage{bm,multicol}
\usepackage[english]{babel}
\usepackage{natbib}
\usepackage{url}
\usepackage[T1]{fontenc}
\usepackage[utf8]{inputenc}
\usepackage[showframe=false]{geometry}
\usepackage{changepage}
\usepackage{authblk}
\usepackage{xcolor}

\newcommand{\newc}{\newcommand}
\newc{\N}{\mbox{N}}
\newc{\1}{\bf{1}}

\begin{document}
\title{\textbf{BFpack}: Flexible Bayes Factor Testing of Scientific Theories in \textsf{R}}

\author{
Joris Mulder, Xin Gu, Anton Olsson-Collentine, Andrew Tomarken, Florian B\"{o}ing-Messing, Herbert Hoijtink, Marlyne Meijerink, Donald R. Williams, Janosch Menke, Jean-Paul Fox, Yves Rosseel, Eric-Jan Wagenmakers, \& Caspar van Lissa}
\date{Pre-print: Nov 16, 2019}

\maketitle

\begin{abstract}
There has been a tremendous methodological development of Bayes factors for hypothesis testing in the social and behavioral sciences, and related fields. This development is due to the flexibility of the Bayes factor for testing multiple hypotheses simultaneously, the ability to test complex hypotheses involving equality as well as order constraints on the parameters of interest, and the interpretability of the outcome as the weight of evidence provided by the data in support of competing scientific theories. The available software tools for Bayesian hypothesis testing are still limited however. In this paper we present a new \textsf{R}-package called \textbf{BFpack} that contains functions for Bayes factor hypothesis testing for the many common testing problems. The software includes novel tools (i) for Bayesian exploratory testing (null vs positive vs negative effects), (ii) for Bayesian confirmatory testing (competing hypotheses with equality and/or order constraints), (iii) for common statistical analyses, such as linear regression, generalized linear models, (multivariate) analysis of (co)variance, correlation analysis, and random intercept models, (iv) using default priors, and (v) while allowing data to contain missing observations that are missing at random.
\end{abstract}



\section{Introduction}
This paper presents a new software package called \textbf{BFpack} which can be used for computing Bayes factors and posterior probabilities for statistical hypotheses in common testing problems in the social and behavioral sciences, medical research, and in related fields. This new package is an answer to the increasing interest of the scientific community to test statistical hypotheses using Bayes factors in the software environment \textsf{R} \citep{Rdevelopment}. Bayes factors enjoy many useful practical and theoretical properties which are not generally shared by classical significance tests. This includes its intuitive interpretation as the relative evidence in the data between two hypotheses, its ability to simultaneously test multiple hypotheses which may contain equality as well as order constraints on the parameters of interest, and its consistent behavior which implies that the true hypothesis will be selected with probability one as the sample size grows. The interested reader is referred to the many important contributions including (but not limited to) \cite{Jeffreys,Berger:1987,Sellke:2001,Wagenmakers:2007,Rouder:2009,Masson:2011,Hoijtink:2011,Wagenmakers:2018,Hoijtink:2019}, and the references therein. This has resulted in an increasing literature where Bayes factors have been used for testing scientific expectations \citep{VanWell:2008,Schoot:2011c,Braeken:2015,Vrinten:2016,vanSchie:2016,MulderWagenmakers:2016,Hoijtink:2017,DeJong:2017,Gronau:2017,Schonbrodt:2017,Wagenmakers:2017b,Raavenswaaij:2019,Flore:2019,Digge:2019,Zondervan:2019}. 

The Bayes factors that are implemented in \textbf{BFpack} are based on recent developments of Bayesian hypothesis testing of location parameters, such as (adjusted) means and regression coefficients \citep{Mulder:2019,Gu:2019,Gu:2017,Mulder:2014b}, variance components, such as group variances and intraclass correlations \citep{BoingMessing:2017,MulderFox:2019}, and measures of association, \citep{Mulder:2016,MulderGelissen:2019}. These Bayes factors can be used for common testing problems in the social and behavioral sciences, and related fields, such as (multivariate) $t$ testing, (multivariate) linear regression, (multivariate) analysis of (co)variance, or correlation analysis. The package allows users to perform (i) exploratory Bayesian tests of whether a model parameter equals zero, is negative, or is positive, and (ii) confirmatory Bayesian tests where users specify a set of competing hypotheses with equality and/or order constraints on the parameters of interest. This will allow users to test their scientific expectations in a direct manner. Thus by providing Bayesian statistical tests for multiple hypotheses with equality as well as order constraints, \textbf{BFpack} makes important contributions to existing software packages, such as \textbf{lmtest} \citep{Hothorn:2019} and \textbf{car} \citep{Fox:2019}, which contain key functions for classical significance tests of a single equality constrained null hypothesis, e.g.,  \texttt{lmtest::coeftest()} and \texttt{car::linearHypothesis()}.

To ensure a simple and user-friendly experience, the different Bayes factors tests are implemented via a single function called \texttt{BF}, which is the workhorse of the package. The function needs a fitted modeling object obtained from a standard \textsf{R} analysis (e.g., \texttt{lm}, \texttt{glm}; see Table \ref{tabletest} for a complete overview), and in the case of a confirmatory test a string that specifies a set of competing hypotheses (example hypotheses are provided in Table \ref{tablehyp}). Another optional argument is the specification of the prior probabilities for the hypotheses. By building on these traditional statistical analyses, which are well-established by the \textsf{R} community, we present users additional statistics measures which cannot be obtained under a frequentist framework, such as default measures of the relative evidence in the data between competing statistical hypotheses as quantified by the Bayes factor.

\begin{table}[h]
\begin{adjustwidth}{-2cm}{}
\caption{\textsf{R} functions, packages, descriptions of tests, parameter of interest, and example name of the parameter that is tested, and the classification whether it is an exact or approximate Bayes factor test. For this table we assume that `\texttt{y1}' is the label of an outcome variable, `\texttt{x1}' is the label of a numeric predictor variable, and `\texttt{g1}' is the label of a level of a grouping (factor) variable. The actual names that are used depend on the names of the variables in the model.}
{\small \setlength\tabcolsep{4pt}\begin{tabular}{llllll}
  \hline
\textsf{R} function & package & test & parameter type & example parameter & Bayes factor \\
\hline
\texttt{t\_test} & \texttt{bain} & Student $t$ test & mean (1-sample test)            & \texttt{mu} & exact\\
               &                  && mean difference  & \texttt{difference} & exact\\
               &&& (2-sample test)\\
\texttt{bartlett\_test} & \texttt{BFpack} & heterogeneity of & group variances & \texttt{g1} & exact\\
&&variances\\
\texttt{aov} & \texttt{stats} & AN(C)OVA & group means & \texttt{g1} & exact\\
\texttt{manova}& \texttt{stats} & MAN(C)OVA & group means & \texttt{g1\_on\_y1} & exact\\
\texttt{lm} & \texttt{stats} & linear regression & regression coefficients & \texttt{x1} & exact\\
&& multivariate regression& regression coefficients & \texttt{x1\_on\_y1} & exact\\
 && & measures of association & \texttt{y1\_with\_y2}, & exact\\
 & & &meas. of assoc. in group & \texttt{y1\_with\_y2\_in\_g1} & exact\\
\texttt{lmer} & \texttt{lme4} & random intercept & group specific & \texttt{g1} & exact\\
& &model &intraclass correlations\\
\texttt{hetcor}& \texttt{polycor} & correlation analysis & measures of association & \texttt{y1\_with\_y2} & approx.\\
\texttt{glm}& \texttt{stats} & generalized linear model & regression coefficients & \texttt{x1} & approx.\\
\texttt{coxph}, & \texttt{survival}& survival analysis & regression coefficients & \texttt{x1} & approx.\\
\texttt{survreg} \\
\texttt{polr}& \texttt{MASS} & ordinal regression & regression coefficients & \texttt{x1} & approx.\\
\texttt{zeroinfl}& \texttt{pscl} & zero-inflated & regression coefficients & \texttt{x1} & approx.\\
&& regression models\\
  \hline
\end{tabular}}
\end{adjustwidth}
\label{tabletest}
\end{table}

\begin{table}[t]
\caption{Example hypothesis tests that can be executed using \textbf{BFpack}.}
{\small \begin{tabular}{ll}
  \hline
 & Example hypotheses\\
  \hline
Exploratory testing & $H_0:\theta=0$ vs $H_1:\theta < 0$ vs $H_2:\theta > 0$.\\
Interval testing & $H_0:|\theta|\le \epsilon$ vs $H_1:|\theta| > \epsilon$, for given $\epsilon>0$.\\
Precise testing & $H_1:\theta_{1}=\theta_{2}=\theta_{3}$ vs $H_2:$ ``not $H_1$''\\
Order testing & $H_1:\theta_{1}>\theta_{2}>\theta_{3}$ vs $H_2:\theta_{1}<\theta_{2}<\theta_{3}$
vs $H_3:$ ``neither $H_1$, nor $H_2$''.\\
Equality and order testing & $H_1:\theta_{12}<\theta_{13}=\theta_{14}$ versus $H_2:$ ``not $H_1$''.\\
\hline
\end{tabular}}
\label{tablehyp}
\end{table}

When testing hypotheses using the Bayes factor, the use of arbitrary or ad hoc priors should generally be avoided \citep{Lindley:1957,Jeffreys,Bartlett:1957,Berger:2001}. Therefore the implemented tests in \textbf{BFpack} are based on default Bayes factor methodology. Default Bayes factors can be computed without requiring external prior knowledge about the magnitude of the parameters. The motivation is that, even in the case prior information is available, formulating informative priors which accurately reflect one's prior beliefs under all separate hypotheses under investigation is a very challenging and time-consuming endeavor \citep{Berger:2006}.

Different default Bayes factors with default priors are implemented for testing different types of parameters, such as location parameters (e.g., means or regression coefficients in univariate/multivariate normal linear models), measures of association (e.g., correlations in multivariate normal distributions), and variance components (e.g., group variances, intraclass correlations). For testing unbounded parameters, such as location parameters and group variances, adjusted fractional Bayes factors \citep{OHagan:1995,Mulder:2014b,BoingMessing:2017} have been implemented. These Bayes factors have analytic expressions and are therefore easy to compute. For testing bounded parameters, such as measures of association and intraclass correlations, proper uniform priors are implemented. When testing intraclass correlations under random intercept models, a novel marginal modeling approach is employed where the random effects are integrated out \citep{MulderFox:2019,Fox:2017,MulderFox:2013}. On the one hand, these tests can be used for testing hypotheses on intraclass correlations based on substantive considerations, and on the other hand, the tests can be used as a tool when building multilevel models as the marginal model approach provides a more general framework for testing covariance structures than regular mixed effects models.

To also facilitate the use of Bayes factors for more general testing problems, an approximate Bayes factor is also implemented which is based on a large sample approximation resulting in an approximate Gaussian posterior distribution. The approximate Bayes factor only requires the (classical) estimates of the parameters that are tested, the corresponding error covariance matrix, and the sample size of the data that was used to get the estimates and covariance matrix. The resulting approximated Bayes factor can be viewed as a Bayesian counterpart of the classical Wald test. This makes the approximate Bayes factor very useful as a general test for statistical hypotheses. Note that even though it is possible to also use the approximate Bayes factor for the testing problems for which exact tailor-made Bayes factors are available in \textbf{BFpack} (e.g., for an \texttt{lm} object), we recommend to use the exact tailored Bayes factors when they are available. The reason is that the exact Bayes factors result in exact quantification of the evidence between statistical hypotheses instead of an approximate quantification of the evidence. Further note that because the sample size is also required to perform the approximate Bayes factor test, \texttt{BF()} cannot be applied directly on the output of classical testing function such as \texttt{lmtest::coeftest()}. In Section \ref{Application5} we show how to do this with one additional step. Table \ref{tabletest} shows for which models an exact Bayes factor is implemented and for which models we make use of the approximation.

Before presenting the statistical methodology and functionality of \textbf{BFpack} it is important to understand what \textbf{BFpack} adds to the currently available software packages for Bayes factor testing. First, the \textsf{R} package \textbf{BayesFactor} \citep{BayesFactor} mainly focuses on precise and interval null hypotheses of single parameters in Student $t$ tests, anova designs, and regression models. It is not designed for testing more complex relationship between multiple parameters. Second, the package \textbf{BIEMS} \citep{Mulder:2012}, which comes with a user interface for \textsf{Windows}, can be used for testing various equality and order hypotheses under the multivariate normal linear model. The computation of the Bayes factors however is too slow for general usage when simultaneously testing many equality constraints as equality constraints are approximated with interval constraints that are made sufficiently small using a computationally intensive step-wise algorithm. Third, the \textbf{bain} package \citep{bain} computes approximated default Bayes factors by assuming normality of the posterior and a default prior. The package has shown good performance for challenging testing problems such as structural equation models. \textbf{BFpack} package also builds on some of the functionality of \textbf{bain} in more complex statistical models. Unlike \textbf{bain} however, the implementation in \textbf{BFpack} builds on existing \textsf{R} functions such as \texttt{dmvnorm} or \texttt{pmvnorm} from the \textbf{mvtnorm} package \citep{Genz:2016} instead of calling external \textsf{Fortran} subroutines. This result in Bayes factors that essentially have zero Monte Carlo errors. Furthermore it is important to note that the Gaussian nature of the default prior in \textbf{bain} may not appropriate when testing bounded parameters, for example, such as measures of association or intraclass correlations, or when the Gaussian approximation of the posterior would be too crude, such as when testing group variances in the case of small sample sizes. Finally the free statistical software environment \textsf{JASP} \citep{JASP2018}, which has contributed tremendously to the use of Bayes factors in psychological research and other research fields, is specifically designed for non-\textsf{R} users by providing a user-friendly graphical user-interface similar to \textsf{SPSS}. The Bayes factors implemented in \textsf{JASP} rely on other packages such as \textbf{BayesFactor} and \textbf{bain}. \textbf{BFpack}, on the other hand, is developed to give \textsf{R} users a flexible tool for testing a very broad class of hypotheses involving equality and/or order constraints on various types of parameters (means, regression coefficients, variance components, and measures of association) under common statistical models by building on standard \textsf{R} functions. 

The paper is organized as follows. Section \ref{sect2} describes the key aspects of the Bayes factor methodology that is implemented in \textbf{BFpack}. This section separately describes Bayes factors for location parameters, for measures of association, and for variance components. Section \ref{sect3} gives a general explanation how the main function \texttt{BF} should be used. Section \ref{sect4} presents 8 different applications of the methodology and software for a variety of testing problems. The paper ends with some concluding remarks in Section \ref{concl}.

\section{Technical background of the default Bayes factors}\label{sect2}
The general form of the hypotheses that can be tested using \textbf{BFpack} consists a set of linear equality constraints and a set of linear order constraints on the vector of model parameters, denoted by $\theta$ of size $P$, i.e.,
\begin{equation}
H_t:\textbf{R}^E\bm\theta=\textbf{r}^E~\&~\textbf{R}^O\bm\theta>\textbf{r}^O,
\label{Ht1}
\end{equation}
where $[\textbf{R}^E|\textbf{r}^E]$ is a $q^E\times P$ augmented matrix specifying the equality constraints and $[\textbf{R}^O|\textbf{r}^O]$ is a $q^O\times P$ augmented matrix specifying the order constraints. A hypothesis index is omitted to keep the notation simple. In the case that $\textbf{R}^{O}$ is of full row rank (which is most often the case), a parameter transformation can be applied according to
\begin{equation}
\left[\begin{array}{c}
\bm\theta^E\\ \bm\theta^O\\ \bm\phi
\end{array}\right]=\textbf{T}\bm\theta=\left[\begin{array}{c}
\textbf{R}^E\\ \textbf{R}^O \\ \textbf{D}
\end{array}\right]\bm\theta,
\end{equation}
where the $q^E$ equality restricted parameters equal $\bm\theta^E=\textbf{R}^E\bm\theta$, the $q^O$ order-restricted parameters equal $\bm\theta^O=\textbf{R}^E\bm\theta$, and the $P-q^O-q^E$ nuisance parameters equal $\bm\phi=\textbf{D}\bm\theta$, where the $(P-q^E-q^O)\times P$ dummy matrix $\textbf{D}$ is chosen such that the transformation is one-to-one. Subsequently the hypothesis can equivalently be formulated as
\begin{equation}
H_t:\bm\theta^E=\textbf{r}^E~\&~\bm\theta^O>\textbf{r}^O,
\label{Ht2}
\end{equation}
where the nuisance parameters $\bm\phi$ are omitted. Note that for most order hypotheses, the matrix $\textbf{R}^{O}$ will be of full row rank. For example, $H_t:\theta_1>\theta_2>\theta_3$ implies that $[\textbf{R}^O|\textbf{r}^O]=\left[\begin{array}{ccc|c}1&-1&0&0 \\ 0&1& -1& 0\end{array}\right]$. Therefore we will work with the formulation in Equation \ref{Ht2} throughout this paper to keep the notation simple. In the case $\textbf{R}^{O}$ is not of full row rank, which is for instance the case for $H_t:(\theta_1,\theta_2)>(\theta_3,\theta_4)$, a similar type of formulation of $H_t$ can be produced as in Equation \ref{Ht2}\footnote{If $\textbf{R}^{O}$ is not of full row rank, then Equation \ref{Ht2} would become $H_t:\bm\theta^E=\textbf{r}^E ~\&~ \tilde{\textbf{R}}^O\bm\theta^O>\tilde{\textbf{r}}^O$, where $\tilde{\textbf{R}}^O=\textbf{R}^O\tilde{\textbf{D}}^{-1}$, where the $(P-q^E)\times P$ matrix $\tilde{\textbf{D}}$ consists of the unique rows of $\textbf{I}_P-\textbf{R}^{E^{\top}}(\textbf{R}^E\textbf{R}^{E^{\top}})^{-1}\textbf{R}^E$, and $\tilde{\textbf{r}}^O=\textbf{r}^O-\textbf{R}^O[\textbf{R}^{E}]^{-1}\textbf{r}^E$, where (generalized) Moore-Penrose inverses are used for the non square matrices.}.

Next we specify a prior for the free (possibly order constrained) parameters under $H_t$, denoted by $\pi_t$, by truncating an unconstrained prior, $\pi_u$, that is specified under an unconstrained alternative model, 
\begin{equation}
\pi_t(\bm\theta^O,\bm\phi) = \pi_u(\bm\theta^O,\bm\phi|\bm\theta^E=\textbf{r}^E) \times \pi_u(\bm\theta^E=\textbf{r}^E)^{-1}\times P(\bm\theta^O>\textbf{r}^O|\bm\theta^E=\textbf{r}^E)^{-1}\times I(\bm\theta^O>\textbf{r}^O),
\label{truncprior}
\end{equation}
where $I(\cdot)$ denotes the indicator function. Using this pair of priors under the constrained hypothesis $H_t$ and the unconstrained alternative hypothesis, we can write the Bayes factor of $H_t$ against $H_u$ as
\begin{equation}
B_{tu} = \frac{\pi_u(\bm\theta^E=\textbf{r}^E|\textbf{Y})}{\pi_u(\bm\theta^E=\textbf{r}^E)}\times \frac{P_u(\bm\theta^O>\textbf{r}^O|\bm\theta^E=\textbf{r}^E,\textbf{Y})}{P_u(\bm\theta^O>\textbf{r}^O|\bm\theta^E=\textbf{r}^E)},
\label{Btu}
\end{equation}
where the first factor is ratio of posterior and prior densities of $\bm\theta$ evaluated at a constant vector $\textbf{R}^E$, which can be viewed as a multivariate Savage-Dickey density ratio \citep{Dickey:1971,Wetzels:2010,Mulder:2010}, and the second factor is a ratio of conditional posterior and prior probabilities that the order constraints hold conditional on the equality constraints. We shall refer to Equation \ref{Btu} as the extended Savage-Dickey density ratio. Different variations have been reported in the literature of this simple expression of the Bayes factor including \cite{Klugkist:2005}, \cite{Pericchi:2008}, \cite{Mulder:2010}, \cite{Gu:2017}, among others. The expression can simply be computed when the marginal and conditional posterior and priors belong to known probability distributions (examples will be given later), and thus direct computation of the marginal likelihood, which can be a challenging problem, can be avoided. The four different statistical measures in Equation \ref{Btu} have the following intuitive interpretations:
\begin{itemize}
\item The marginal posterior density evaluated at $\bm\theta^E=\textbf{r}^E$ (numerator of first factor) is a measure of the \textit{relative fit of the equality constraints} of $H_t$ relative to $H_u$ as a large (small) posterior value under the unconstrained model indicates that there is evidence in the data that $\bm\theta^E$ is (not) close to $\textbf{r}^E$.
\item The conditional posterior probability of $\bm\theta^O>\textbf{r}^O$ given $\bm\theta^E=\textbf{r}^E$ (numerator of second factor) is a measure of the \textit{relative fit of the order constraints} of $H_t$ relative to $H_u$ as a large (small) probability under the unconstrained model indicates that there is evidence in the data that the order constraints (do not) hold.
\item The marginal prior density evaluated at $\bm\theta^E=\textbf{r}^E$ (denominator of first factor) is a measure of the \textit{relative complexity of the equality constraints} of $H_t$ relative to $H_u$ as a large (small) prior value indicates that the prior for $\bm\theta^E$ is (not) concentrated around $\textbf{r}^E$, and thus there is little (big) difference between the precise formulation $\bm\theta^E=\textbf{r}^E$ or the unconstrained formulation $H_u$.
\item The conditional prior probability of $\bm\theta^O>\textbf{r}^O$ given $\bm\theta^E=\textbf{r}^E$ (denominator of second factor) is a measure of the \textit{relative complexity of the order constraints} of $H_t$ relative to $H_u$ as a large (small) probability under the unconstrained model indicates that the order constrained subspace under $H_t$ is relatively large (small), indicating that the constrained model is complex (simple).
\end{itemize}

It is important to note that by conditioning on $\bm\theta^E=\textbf{r}^E$ in Equation \ref{truncprior}, we make specific assumptions about the prior of the free parameters under $H_t$ in relation to the unconstrained prior \citep{Verdinelli:1995,Marin:2010}, and therefore the expression should be used with some care \citep[see also][for an interesting discussion on this topic]{Consonni:2008}. Below we provide examples of Bayes factors that can and Bayes factors that cannot be expressed as an extended Savage-Dickey density ratio.

%
%
%
%
%
%
%

\subsection{Testing location parameters}
Many common testing problems in statistical science involve testing of location parameters that determine the `location' or `shift' of the distribution of the data. Examples of location parameters are means, regression coefficients, or factor loadings. These parameters are unbounded for which flat improper priors are specified under an objective Bayesian estimation framework, i.e., $\pi_u(\bm\theta)\propto 1$.

Fractional Bayes methodology is an effective framework for testing location parameters. Informative (subjective) prior specification is avoided by splitting the data in a minimal fraction that is used for updating a noninformative improper prior to a proper default prior and a maximal fraction that is used for hypothesis testing \citep{OHagan:1995,DeSantis:1999}. Despite the various useful properties of fractional Bayes factors \citep[e.g., consistency, coherence when testing multiple hypotheses, invariance to transformations of the data,][]{OHagan:1997}, an adjustment was needed in order for the fractional Bayes factor to function as an Occam's razor when testing order hypotheses \citep{Mulder:2014b,MulderOlsson:2019}. This is achieved by shifting the default prior to the boundary of the constrained space\footnote{When testing a constrained hypothesis of the form of Equation \ref{Ht2}, the default prior is centered on the boundary which implies that the prior is centered around $\bm\theta_0$ with $\textbf{R}^E\bm\theta_0=\textbf{r}^E$ and $\textbf{R}^O\bm\theta_0=\textbf{r}^O$.}. In the simple case when testing $\theta<0$ versus $\theta>0$, the default prior would be centered at 0 (instead of around the MLE) so that the prior probabilities of $\theta<0$ and $\theta>0$ under the unconstrained model are equal to 0.5, which suggests that a negative effect is equally likely as a positive effect. Centering the unconstrained prior to the boundary also resulted in desirable testing behavior of order hypotheses when using intrinsic Bayes factors \citep{Mulder:2010,Mulder:2014a} and when using the BIC \citep{MulderRaftery:2019}

Interestingly when testing location parameters with flat improper priors, the adjusted fractional Bayes factor (and the fractional Bayes factor as well) of $H_t$ against $H_u$ can be expressed as an extended Savage-Dickey density ratio as in Equation \ref{Ht2}, i.e.,
\begin{equation}
B_{tu}^F = \frac{\pi_u(\bm\theta^E=\textbf{r}^E|\textbf{Y})}{\pi_u^*(\bm\theta^E=\textbf{r}^E|\textbf{Y}^b)}\times \frac{P_u(\bm\theta^O>\textbf{r}^O|\bm\theta^E=\textbf{r}^E,\textbf{Y})}{P_u^*(\bm\theta^O>\textbf{r}^O|\bm\theta^E=\textbf{r}^E,\textbf{Y}^b)},
\label{FBFtu}
\end{equation}
where the distributions conditional on $\textbf{Y}^b$ in the denominator denote the unconstrained default priors that contain a minimal fraction $b$ of the complete data $\textbf{Y}$, and the asterisk ($^*$) denotes the default prior adjustment. When the data contains information from different groups and the sample sizes highly varies across groups, it is generally recommended to use group specific fractions to properly control the amount of prior information from each group \citep{DeSantis:2001,Hoijtink:2018}.

\subsubsection{Univariate/multivariate normal linear models}
Recently, \citep{Mulder:2019} derived the adjusted fractional Bayes factor for testing hypotheses under the multivariate normal linear model with multiple groups. Under this model the unconstrained posterior of the matrix of location parameters follows a matrix Student $t$ distribution, and the unconstrained default prior has a matrix Cauchy distribution, i.e.,
\begin{equation}
B_{tu}^F = \frac{\mathcal{T}(\bm\theta^E=\textbf{r}^E|\textbf{Y})}{\mathcal{C}(\bm\theta^E=\textbf{r}^E|\textbf{Y}^{\textbf{b}})}\times \frac{\mathcal{T}(\bm\theta^O>\textbf{r}^O|\bm\theta^E=\textbf{r}^E,\textbf{Y})}{\mathcal{C}(\bm\theta^O>\textbf{r}^O|\bm\theta^E=\textbf{r}^E,\textbf{Y}^{\textbf{b}})},
\label{FBFtuMVN}
\end{equation}
where the $\mathcal{T}_u$ and $\mathcal{C}_u$ denote the unconstrained matrix Student $t$ and matrix Cauchy distribution, respectively, and $\textbf{b}$ denotes a vector of minimal fractions that are group specific.

Under these matrix-variate distributions, the posterior and prior densities at $\textbf{r}^E$, and the conditional posterior and prior probabilities that the order constraints hold do not have analytic expressions. In \textbf{BFpack}, these quantities are computed using Monte Carlo integration. We use the fact that draws from a matrix Student $t$ and matrix Cauchy distribution can be obtained by first sampling a covariance matrix from an inverse Wishart distribution, and subsequently drawing the matrix of location parameters from its respective matrix Gaussian distribution conditional on the drawn covariance matrix \citep{BoxTiao}. Therefore, the posterior density evaluated at $\bm\theta^E=\textbf{r}^E$ can be obtained by repeatedly drawing covariance matrices from its marginal posterior, and subsequently, computing the posterior density as the arithmetic average of the Gaussian densities evaluated at $\bm\theta^E=\textbf{r}^E$, which have analytic expressions. The Gaussian densities are computed using the \texttt{dmvnorm} function from the \textbf{mvtnorm} package \citep{Genz:2016}. Such a procedure is also implemented to obtain the prior density, and the conditional prior and posterior probabilities. The Gaussian probabilities are obtained using the \texttt{pmvnorm} function from the \textbf{mvtnorm} package.

In case the constraints are formulated only on the effects belonging to the same dependent variable, or only on the effects belonging to the same independent (predictor) variable, the marginal and conditional distributions for the unconstrained parameters follow multivariate Student $t$ distributions. The respective measures of relative complexity and fit then have analytic expressions which are efficiently computed using \texttt{dmvt} and \texttt{pmvt} (\textbf{mvtnorm} package) in \textbf{BFpack}. Finally note that fractional Bayes factors between the constrained hypotheses using the coherence property of the Bayes factor, i.e., $B_{12}=B_{1u}/B_{2u}$.

This Bayes factor test is executed when the data are fitted using the \textsf{R} functions \texttt{t\_test}, \texttt{lm}, \texttt{aov}, and \texttt{manova}. Note that the usual $t$ test function in \textsf{R}, \texttt{t.test}, cannot be used because the output (of class \texttt{htest}) does not contain the observed sample means and sample variances of the two groups which are needed for the computation of the Bayes factors. For this reason, the equivalent function \texttt{t\_test} was used (from the \textbf{bain} package) which contains the sample means and variances in addition to the standard output of \texttt{t.test}.

\subsubsection{General statistical models}
Under more complex statistical models where the four quantities in Equation \ref{FBFtu} do not have analytic expressions or when they cannot be computed efficiently via Monte Carlo estimation, an approximation of the adjusted fractional Bayes factor can be used \citep{Gu:2017,Gu:2019}. This approximation relies on large sample theory where the unconstrained posterior and default prior are approximated with Gaussian distributions. As such this approximate default Bayes factor can be viewed as a Bayesian counterpart of the classical Wald test. It can be used when no exact Bayes factor is available (see the sixth column in Table \ref{tabletest}).

First the nuisance parameters are integrated out to yield the marginal posterior for $(\bm\theta^E,\bm\theta^O)$. Following large sample theory \citep[][Ch. 4]{Gelman:2014}, this posterior can then be approximated with a multivariate Gaussian distribution using the MLE and error covariance matrix. The approximated Gaussian posteriors can then be used to get estimates of the posterior quantities in the numerators in Equation \ref{FBFtu}. The corresponding default prior for $(\bm\theta^E,\bm\theta^O)$ is obtained by raising the posterior to a minimal fraction $b$, which results in a multivariate Gaussian distribution where the error covariance matrix is multiplied with the reciprocal of the minimal fraction, and the mean is shifted towards the boundary of the constrained space. This default Bayes factor can then be written as
\begin{equation}
B^F_{tu} = \frac{\mathcal{N}(\bm\theta^E=\textbf{r}^E|\textbf{Y})}{\mathcal{N}(\bm\theta^E=\textbf{r}^E|\textbf{Y}^b)}\times \frac{\mathcal{N}(\bm\theta^O>\textbf{r}^O|\bm\theta^E=\textbf{r}^E,\textbf{Y})}{\mathcal{N}(\bm\theta^O>\textbf{r}^O|\bm\theta^E=\textbf{r}^E,\textbf{Y}^b)},
\label{FBFtuBain}
\end{equation}
where $\mathcal{N}$ denotes an unconstrained multivariate (or matrix-variate) normal distribution. In \textbf{BFpack} the posterior and prior densities, and posterior and prior probabilities are directly computed using \texttt{dmvnorm} and \texttt{pmvnorm} functions from the \textbf{mvtnorm} package, respectively. In the case the matrix of order constraints is not of full row rank, \texttt{pmvnorm} cannot be used for computing the needed probabilities. In this special case the \texttt{bain} function is called from the \textbf{bain} package.

Hence, in order to compute the approximated Bayes factor in Equation \ref{FBFtuBain} only the MLEs, the error covariance matrix, and the sample size are needed. The minimal fraction is then set equal to the number of constraints that are tested divided by the sample size \citep{Gu:2019}. These three elements can simply be extracted from fitted model objects obtained using other packages in \textsf{R}. Currently \textbf{BFpack} supports objects of class \texttt{glm}, \texttt{coxph}, \texttt{polr}, \texttt{survreg}, and \texttt{zeroinfl}. When executing \texttt{BF()} on an object of these classes, the function \texttt{BF.default} is called which extracts the estimates, the error covariance matrix, and the sample size from the fitted model object to compute Equation \ref{FBFtuBain} for the hypotheses of interest. This Bayes factor is also called when the input argument for \texttt{BF()} is a named vector with the estimates, its corresponding error covariance matrix, and the sample size, together with the hypotheses of interest. This is explained in more detail in Section \ref{sect3}. 


\subsection{Testing measures of association}
Correlation coefficients and other measures of association play a central role in applied research to quantify the strength of the linear relationship between two variables, possibly controlling for other variables. Measures of association abide two conditions. First they are bounded between -1 and 1, and second they lie in a correlation matrix which must be positive definite. The second condition implies that a correlations need to satisfy a complex set of constraints \citep[e.g.,][]{Rousseeuw:1994}. The volume of this subspace for increasing dimensions of the correlation matrix was explored by \cite{Joe:2006}.

As measures of association are bounded, fractional Bayes methodology is not needed as the noninformative joint uniform prior for the correlations in the correlation matrix is already proper, and thus a regular default Bayes factor can be computed. This was also recommended by \cite{Jeffreys:1935}. This joint uniform prior assumes that any configuration of correlations that results in a positive definite correlation matrix is equally likely a priori. Equivalently, proper uniform priors can be formulated for the measures of association under the constrained hypotheses under investigation. It is easy to show that this proper uniform prior under $H_t$ can be written as a truncation of the unconstrained joint uniform prior as in Equation \ref{truncprior}, and therefore, the Bayes factor of constrained hypothesis against an unconstrained alternative can be expressed as an extended Savage-Dickey density ratio in Equation \ref{Btu}, where the unconstrained prior in the denominator is the joint uniform prior and the unconstrained posterior is proportional to the likelihood and this uniform prior \citep{MulderGelissen:2019}. Furthermore as was shown by \cite{Mulder:2016} the unconstrained posterior for the measures of association can be well approximated with a multivariate normal distribution after a Fisher transformation of the parameters. This can be explained by the fact that the sample correlation and the population correlation have a similar role in the likelihood \citep{Johnson:1970}, and therefore approximate normality is achieved for the posterior when using a noninformative prior such as the employed joint uniform prior. The Bayes factor on measures of association that is implement in \textbf{BFpack} can therefore be written as\footnote{Note there is a slight abuse of notation in Equation \ref{BFtuCor} as both the numerator and denominator for $\bm\theta$ have to lie on the same scale to avoid the Borel-Kolmogorov paradox \citep{Wetzels:2010}. In the computation in \textbf{BFpack}, the numerator and denominator are either both computed under the Fisher transformed space or under the untransformed space depending on the test.}
\begin{equation}
B_{tu} = \frac{\mathcal{N}(\bm\theta^E=\textbf{r}^E|\textbf{Y})}{\mathcal{U}(\bm\theta^E=\textbf{r}^E)}\times \frac{\mathcal{N}(\bm\theta^O>\textbf{r}^O|\bm\theta^E=\textbf{r}^E,\textbf{Y})}{\mathcal{U}(\bm\theta^O>\textbf{r}^O|\bm\theta^E=\textbf{r}^E)},
\label{BFtuCor}
\end{equation}
To obtain the prior measures for relative complexity, numerical estimates can be obtained by approximating the joint prior using unconstrained draws, from which the prior density and probability can simply be computed using the number of draws satisfying the constraints. In \textbf{BFpack} this is done by calling \textsf{Fortran 90} subroutines from \textsf{R}. 

This Bayes factor test can be executed when the fitted model is a multivariate linear regression model (so that the fitted object is of class \texttt{mlm}). 
Furthermore, an approximation of the Bayes factor is obtained when the fitted model object is obtained using the \textsf{R} function \texttt{hetcor} \citep[from the \textbf{polycor} package;][]{polycor:2016}. The mean vector and covariance matrix of the approximately multivariate normal posterior for the measures of association are obtained by extracting the estimates and standard errors from the \texttt{hetcor} object.

%

\subsection{Testing variance components}\label{SectionVariances}

\subsubsection{Testing group variances}
Testing the heterogeneity of group variances plays a central role in psychological science and related fields. A default Bayes factor for testing equality and order hypotheses was developed by \citep{BoingMessing:2017} using adjusted fractional Bayes methodology. As variance parameters belong to the family of scale parameters, a scale adjustment is needed to obtain a default Bayes factor that functions as an Occam's razor for order hypotheses on variances \citep[instead of a location shift as for location parameters, see][]{BoeingMessing:2016,BoeingMessing:2018}. Because the noninformative independence Jeffreys prior for group variances across competing equality constrained hypotheses does not satisfy Equation \ref{truncprior}, the fractional Bayes factor for the equality part (i.e., the first factor in Equation \ref{Ht2}) cannot be expressed as a Savage-Dickey density ratio but the ratio of (conditional) probabilities is present. The Bayes factor for the group variance test can be written as follows
\begin{equation}
B^F_{tu} = B^F_{t'u} \times \frac{\mathcal{IG}(\bm\theta^O>\textbf{r}^O|\bm\theta^E=\textbf{r}^E,\textbf{Y})}{\mathcal{IG}(\bm\theta^O>\textbf{r}^O|\bm\theta^E=\textbf{r}^E,\textbf{Y}^b)},
\label{FBFtuVar}
\end{equation}
where $B^F_{t'u}$ denotes the fractional Bayes factor of hypothesis $H_{t'}:\bm\theta^E=\textbf{r}^E$ \citep[i.e., hypothesis $H_t$ where the order constraints are omitted, see also][]{Pericchi:2008} against $H_u$, and $\mathcal{IG}$ denotes an unconstrained inverse gamma distribution. We refer the interested reader to \citep{BoingMessing:2017} for the mathematical expressions and derivations.

This Bayes factor test can be executed when the fitted model is obtained from the \textsf{R} function \texttt{bartlett\_test}, designed for \textbf{BFpack}. This test is equivalent to the usual \texttt{bartlett.test} but the output object (of class \texttt{BF\_bartlett}) also contains sample variances and sample sizes which are needed for computing the Bayes factors in Equation \ref{FBFtuVar}.

\subsubsection{Testing between-cluster variances and intraclass correlations in mixed effects models}
The multilevel or mixed effects model is the gold standard for modeling hierarchically structured data. In the mixed effects model the within-clusters variability is separately modeled from the between-clusters variability. The intraclass correlation plays a central role as a measure of the relative degree of clustering in the data where an intraclass correlation close to 1 (0) indicates a very high (low) degree of clustering in the data.

Despite the widespread usage of mixed effects models in the (applied) statistical literature, there are few statistical tests for testing variance components; exceptions include \cite{Westfall:1996,Garcia:2007,Saville:2009,Thalmann:2017}. The complicating factor is that testing whether the between-cluster variance equals zero is a boundary problem. In \textbf{BFpack} a Bayes factor testing procedure is implemented for intraclass correlations (and random intercept variances) under a marginal modeling framework where the random effects are integrated out \citep{MulderFox:2019,Fox:2017,MulderFox:2013}. Under the marginal model the intraclass correlations become covariance parameters which may attain negative values. This crucial step allows us to test the appropriateness of a random effects model using the posterior probability that an intraclass correlation is positive. The implemented Bayes factors make use of stretched uniform priors for the intraclass correlations in the interval $(-\frac{1}{p-1},1)$, where $p$ is the cluster size. This prior is equivalent to a shifted-$F$ prior on the between-cluster variances. Similar as when testing group variances, the equality part of the Bayes factor of a constrained hypothesis on the intraclass correlations against an unconstrained alternative cannot be expressed as a Savage-Dickey density ratio. The Bayes factor can be written as
\begin{equation}
B_{tu} = B_{t'u} \times \frac{\text{shifted-}\mathcal{F}(\bm\theta^O>\textbf{r}^O|\bm\theta^E=\textbf{r}^E,\textbf{Y})}{\text{shifted-}\mathcal{F}(\bm\theta^O>\textbf{r}^O|\bm\theta^E=\textbf{r}^E)},
\label{BFtuICC}
\end{equation}
where $\text{shifted-}\mathcal{F}$ refers to the fact that the conditional draws for the between cluster variances are drawn from shifted-$F$ priors in the Gibbs sampler. The marginal likelihood is estimated using importance sampling; see \cite{MulderFox:2019} for the mathematical details.

These Bayes factors can be used for testing the degree of clustering in the data (e.g., testing whether clustering is present among students from different schools), or for testing whether the degree of clustering varies across different cluster categories (e.g., testing the degree of clustering among students from private schools against the degree of clustering among students from public schools). To execute these tests, an object from the \texttt{lmer} function with random intercepts (which may be category specific) is needed. Currently \textbf{BFpack} only supports intraclass correlation testing in the case of equally sized clusters.

\subsection{Bayes factor computation for data with missing observations}
Bayesian (and non-Bayesian) hypothesis testing in the case the data contains missing observations has not received a lot of attention in the literature. This is quite surprising as missing data are ubiquitous in statistical practice. If the data contain missing observations, listwise deletion is generally not recommended as this results in a loss of information and possible bias \citep{Rubin:1987,Rubin:1996}. Multiple imputation is generally the recommended method in which many imputed data sets are randomly created using an imputation model. The analyses are then performed over all the imputed data sets, and averaged in a proper manner \citep{LittleRubin2002}.

In the case of model uncertainty, properly handling missing data may become increasingly complex as different imputation models need to be used for computing the marginal likelihoods under the different hypotheses. \cite{Hoijtink:2018b} however showed that the computation can be considerably simplified for specific Bayes factors and testing problems. This is also the case for Bayes factors that can be expressed as the extended Savage-Dickey density ratio in Equation \ref{Btu}. The reason is that the four key quantities (i.e., the measures of relative fit and relative complexity for the equality and order constraints) are all computed under the same unconstrained model. Therefore we only need to get an unbiased estimate of the unconstrained posterior (and possibly unconstrained default prior in the case of a data-based prior), and use this to estimate the four key quantities. If we write a complete data matrix $\textbf{Y}$ as a data matrix which only contains the observations, $\textbf{Y}^o$, and a data matrix which only contain the missings as $\textbf{Y}^m$, the relative fit of the equality constraints can be computed as
\begin{eqnarray*}
\pi_u(\bm\theta^E=\textbf{r}^E|\textbf{Y}^o) &=& \int \pi_u(\bm\theta^E=\textbf{r}^E|\textbf{Y}^o,\textbf{Y}^m) \pi_u(\textbf{Y}^m|\textbf{Y}^o) d\textbf{Y}^m\\
&\approx& M^{-1}\sum_{m=1}^M \pi_u(\bm\theta^E=\textbf{r}^E|\textbf{Y}^o,\textbf{Y}^{(m)}),
\end{eqnarray*}
where $\textbf{Y}^{(m)}$ is the $m$-th set of imputed missing observations given the observed data matrix $\textbf{Y}^o$, for $m=1,\ldots,M$. Similar expressions can be obtained for the other three measures. Section \ref{SectionMissing} illustrates how to compute Bayes factors and posterior probabilities via the output from \textbf{BFpack} in the presence of missing data using the imputation software of the \textbf{mice} package \citep{mice:2019}.

\section{Bayes factor testing using the package}\label{sect3}
The Bayes factors described in the previous section can be executed by calling the function \texttt{BF}. The function has the following arguments:
\begin{itemize}
\item \texttt{x}, a fitted model object that is obtained using a \textsf{R}-function. An overview \textsf{R}-functions that are currently supported can be found in Table \ref{tabletest}.
\item \texttt{hypothesis}, a string that specifies the hypotheses with equality and/or order constraints on the parameters of interest.
\begin{itemize}
\item The parameter names are based on the names of the estimated effects. Thus, if the coefficients in a fitted \texttt{lm} object have the names \texttt{weight}, \texttt{height}, and \texttt{length}, then the constraints in the \texttt{hypothesis} argument should be formulated on these names.
\item Constraints within a hypothesis are separated with an ampersand `\texttt{\&}'. Hypotheses are separated using a semi-colon `\texttt{;}'. For example \texttt{hypothesis = "weight > height }\texttt{\&}\texttt{ height > 0; weight = height = 0"} implies that the first hypothesis assumes that the effect of \texttt{weight} is larger than the effect of \texttt{height} and that the effects of \texttt{height} is positive, and the second hypothesis assumes that the two effects are equal to zero. Note that the first hypothesis could equivalently have been written as \texttt{weight > height > 0}.
\item Brackets, `\texttt{(}' and `\texttt{)}', can be used to combine constraints of multiple hypotheses. For example \texttt{hypothesis = "(weight, height, length) > 0"} denotes a hypothesis where both the effects of \texttt{weight}, \texttt{height}, and \texttt{length} are positive. This could equivalently have been written as \texttt{hypothesis = "weight > 0 }\texttt{\&}\texttt{ height > 0 \texttt{\&} length > 0"}.
\item In the case the subspaces under the hypotheses do not cover the complete parameter space, a complement hypothesis is automatically added. For example, if an equality hypothesis and an order hypothesis are formulated, say, \texttt{hypothesis = "weight} \texttt{= height = length; weight > height >} \texttt{length"}, the \texttt{complement} hypothesis covers the remaining subspace where neither "\texttt{weight = height = }\\ \texttt{length}" holds, nor \texttt{"weight >} \texttt{height > length"} holds.
\item In general we recommended not to specify order hypotheses that are nested, such as \texttt{hypothesis = "weight > height > length; weight > (height, }\texttt{length)"},\\ where the first hypothesis (which assumes that the effect of \texttt{weight} is larger than the effect of \texttt{height}, and the effect of \texttt{height} is larger than the effect of \texttt{length}) is nested in the second hypothesis (which assumes that the effects of \texttt{weight} is largest but no constraints are specified between the effects of \texttt{height} and \texttt{length}). The reason is that the Bayes factor for the simpler hypothesis against the more complex hypothesis would be bounded. Therefore the scale of the Bayes factor would become more difficult to interpret, and the evidence could not accumulate to infinity for the true hypothesis if the true hypothesis would be the smaller order hypotheses \citep[e.g., see][]{Mulder:2010}. If however a researcher has theoretical reasons to formulate nested order hypotheses these can be formulated and tested using the \texttt{BF} function of course.
\item The default setting is \texttt{hypothesis = NULL}, which only gives the output for exploratory tests of whether each parameter is zero, negative, or positive when assuming equal prior probabilities, e.g., \texttt{hypothesis = "weight = 0; weight < 0; weight > 0}, for the effect of \texttt{weight}. This exploratory tests is also executed when a confirmatory test is of interest via the \texttt{hypothesis} argument.
\item When testing hypotheses on variance components (Section \ref{SectionVariances}), only simple constraints are allowed where a parameter is equal to, greater than, or smaller than another parameter. When testing intraclass correlations, the intraclass correlation can also be compared to 0 under a hypothesis.
\end{itemize}
\item \texttt{prior}, a numeric vector of prior probabilities of the hypotheses. The default setting is \texttt{prior = NULL} which specifies equal prior probabilities.
\end{itemize}
In the case the class of the fitted model \texttt{x} is not supported, \texttt{BF.default()} is called which executes the approximate Bayes factor (sixth column of Table \ref{tabletest}). In this case, the following (additional) arguments are required:
\begin{itemize}
\item \texttt{x}, a named numeric vector of the estimates (e.g., MLE) of the parameters of interest where the labels are equal to the names of the parameters which are used for the \texttt{hypothesis} argument.
\item \texttt{Sigma}, the approximate posterior covariance matrix (e.g,. error covariance matrix) of the parameters of interest.
\item \texttt{n}, the sample size that was used to acquire the estimates and covariance matrix.
\end{itemize}
The output is an object of class \texttt{BF}. When printing an object of class \texttt{BF} via the \texttt{print()} function, the posterior probabilities for the hypotheses under evaluation are provided, or, in the case \texttt{hypothesis = NULL}, the posterior probabilities are given for exploratory tests of whether each parameter is zero, negative, or positive. The \texttt{summary()} function shows the results for the exploratory tests, and if hypotheses are specified in the \texttt{hypothesis} argument, the results of the confirmatory tests consisting of the posterior probabilities of the hypotheses of interest, the evidence matrix which shows the Bayes factor between each pair of two hypotheses, a specification table which shows all the measures of relative fit and complexity for the equality and/or order constraints of the hypotheses, and an overview of the hypotheses that are tested.

\section{Applications}\label{sect4}
This section presents a variety of testing problems that can be executed using \textbf{BFpack}. On \underline{https://github.com/cjvanlissa/BFpack\_paper}, a \textsf{R Markdown} version of the paper can be found to facilitate the reproducibility of the analyses.

\subsection{Application 1: Bayesian $t$ testing in medical research}
The example for a one-sample $t$ test was discussed in \cite[][p. 196]{Howell:2012}, and originally presented in \cite{Rosa:1998}. An experiment was conducted to investigate whether practitioners of the therapeutic touch (a widely used nursing practice) can effectively identify which of their hands is below the experimenter's under blinded condition. Twenty-eight practitioners were involved and tested 10 times in the experiment. Researchers expected an average of 5 correct answers from each practitioner as it is the number by chance if they do not outperform others. In this example, the data are the number of correct answers from 0 to 10 of $n=28$ practitioners. The null and alternative hypotheses are $H_1:\mu=5$ and $H_2:\mu > 5$ where $\mu$ is the mean of the data. If $H_1:\mu=5$ is true, it means that practitioners give correct answers by chance, whereas if $H_2:\mu > 5$, this implies that practitioners do better than expected by random chance. The \texttt{BF} function automatically adds the complement hypothesis, $H_3:\mu < 5$, which would imply that practitioners do worse than expected by chance. As there is virtually no prior belief that $H_3$ may be true, and we (for this example) assume that the hypotheses of interest, $H_1$ and $H_2$, are equally likely a priori we set the prior probabilities for $H_1$, $H_2$, and $H_3$ in the confirmatory test to 0.5, 0.5, and 0, respectively, using the \texttt{prior} argument.

Hypotheses $H_1:\mu=5$ versus $H_2:\mu> 5$ are tested used the frequentist $t$ test function \texttt{t\_test} from the \textsf{R} package \texttt{bain} and Bayesian $t$ test function \texttt{BF} in the \textsf{R} package \textbf{BFpack}.
\begin{verbatim}
R> install.packages("BFpack")
R> library(BFpack)
R> install.packages("bain")
R> library(bain)
R> ttest1 <- t_test(therapeutic, alternative = "greater", mu = 5)
R> print(ttest1)
R> BF1 <- BF(ttest1, hypothesis = "mu = 5; mu > 5", prior = c(.5,.5,0))
R> summary(BF1)
\end{verbatim}
The first six lines install and load the \textsf{R} package \textbf{BFpack}.
In the 8th line, \texttt{t\_test} function renders classical right one-sided $t$ test and stores the result in object \texttt{ttest1}, which contains $t$ value, degree of freedom, and $p$ value, as well as $95\%$ confidence interval: \\
\begin{verbatim}
data:  therapeutic
t = -1.9318, df = 27, p-value = 0.968
alternative hypothesis: true mean is greater than 5
95 percent confidence interval:
 3.857523      Inf
sample estimates:
mean of x 
 4.392857 
\end{verbatim}
The $p$ value indicates that there is no reason to reject $H_1$ against the right one-tailed alternative using traditional choices for the significance level. This however does not imply there is evidence in favor of $H_1$ as significance tests cannot be used for quantifying evidence for a null hypothesis. Next, the object \texttt{ttest1} is used as the input for the \texttt{BF} function for a Bayesian $t$ test in the 11th and 13th line. Specifically, \texttt{BF(ttest1, hypothesis="mu = 5; mu > 5", prior = c(.5,.5,0))} executes a confirmatory Bayes factor test where the hypothesis $H_1: \mu=5$ is tested against $H_2: \mu> 5$. The complement hypothesis, $H_3: \mu< 5$ in this case, is automatically added. The argument \texttt{prior=c(.5,.5,0)} indicates that the first two hypotheses are assumed to be equally likely a priori and the third hypothesis does not receive any prior support. The \texttt{summary(BF1)} on line 12 shows all the results of the exploratory and confirmatory tests:
\begin{verbatim}
Call:
BF.t_test(x = ttest1, hypothesis = "mu=5")

Bayesian hypothesis test
Type: Exploratory
Object: t_test
Parameter: means
Method: generalized adjusted fractional Bayes factor

Posterior probabilities:
   Pr(=5) Pr(<5) Pr(>5)
mu  0.345  0.634  0.021

Bayesian hypothesis test
Type: Confirmatory
Object: t_test
Parameter: means
Method: generalized adjusted fractional Bayes factor

Posterior probabilities:
   Pr(hypothesis|data)
H1               0.943
H2               0.057
H3               0.000

Evidence matrix:
      H1     H2    H3
H1 1.000 16.473 0.544
H2 0.061  1.000 0.033
H3 1.838 30.276 1.000

Specification table:
   comp_E comp_O fit_E fit_O  BF_E  BF_O    BF   PHP
H1  0.195    1.0 0.205 1.000 1.053 1.000 1.053 0.943
H2  1.000    0.5 1.000 0.032 1.000 0.064 0.064 0.057
H3  1.000    0.5 1.000 0.968 1.000 1.936 1.936 0.000

Hypotheses:
H1: mu=5
H2: mu>5
H3: complement
\end{verbatim}
The results of the exploratory tests show that the posterior probabilities of the precise null ($\mu=5$), a negative effect ($\mu<5$), and a positive effect ($\mu>5$) are 0.345, 0.634, and 0.021, respectively, while assuming equal prior probabilities for the three hypotheses, i.e., $P(H_1)=P(H_2)=P(H_3)=\frac{1}{3}$. The results from the exploratory test show that the presence of a negative is most plausible given the observed data but the evidence is relatively small as there is still a probability of 0.345 that the precise null is true, and a small probability of 0.021 that there is a positive population effect.

The exploratory test however ignores the researchers prior expectations that the first two hypotheses were assumed to be equally likely while there was no reason to believe that the third hypothesis could be true, i.e., $P(H_1)=P(H_2)=\frac{1}{2}$ and $P(H_3)=0$. Taking these prior probabilities into account, the confirmatory test shows that there is clearly most evidence that a therapeutic touch does not exist ($H_1$) with a posterior probability of 0.943, followed by the hypothesis that a therapeutic touch exists ($H_2$) with a posterior probability of 0.057. Furthermore the \texttt{Evidence matrix} shows that the Bayes factor for $H_1$ against $H_2$ equals 16.473, which is equal to the ratio of the (non rounded) posterior probabilities of the respective hypotheses as equal prior probabilities were specified.

Finally the \texttt{Specification table} shows that the measures of relative complexity and relative fit for the constrained hypotheses. The relative fit of the one-sided hypotheses ($H_2:\mu>5$ and $H_3:\mu<5$) equal 0.5 (column \texttt{comp\_O}), which can be explained by the fact that the implied one-sided subspaces cover half of the unconstrained space. Furthermore the posterior probability mass in the region $\mu>5$ and $\mu<5$ under the unconstrained model equal 0.032 and 0.968 (column \texttt{fit\_O}), respectively, which quantify the relative fit of the one-sided hypotheses. The unconstrained default prior and posterior density at $\mu=5$ equal 0.195 and 0.205 (column \texttt{comp\_E} and \texttt{fit\_E}), which quantify the relative complexity and fit of the precise hypothesis, respectively.

\subsection{Application 2: 2-way ANOVA to investigate numerical judgement}\label{Application2}
\cite{Janiszewski:2008} executed several experiments to investigate the numerical judgments of participants. In one of the experiments (referred to as `4a') the outcome variable was the amount by which the price for a television estimated by a participant differed from an anchor price (expressed by means of a $z$ score), and the two factors where (1) whether the anchor price was rounded, e.g., \$5000, or precise, e.g., \$4989 (\texttt{anchor} = \texttt{rounded} or \texttt{precise}, respectively); and (2) whether the participants received a suggestion that the estimated price is close to the anchor value or whether they did not receive this suggestion (\texttt{motivation} = \texttt{low} or \texttt{high}, respectively). An example of a question, with \texttt{anchor} = \texttt{rounded} and \texttt{motivation} = \texttt{low}, was: ``The retail price of a TV is \$5000 (rounded). The actual price is only slightly lower than the retail price. Can you guess the price?''. Alternatively, by changing `\$5000' to `\$4989' in the question a \texttt{precise} anchor price is obtained. By changing `slightly lower' to `lower' a question with a \texttt{high} motivation is obtained.

This $2 \times 2$ ANOVA design can be tested using \textbf{BFpack} as follows
\begin{verbatim}
R> aov1 <- aov(price ~ anchor * motivation, data = tvprices)
R> BF(aov1)
\end{verbatim}
For an object of class \texttt{aov}, \textbf{BFpack} also provides the Bayes factors for the existence of the main effects and interactions effects in the exploratory tests
\begin{verbatim}
Bayesian hypothesis test
Type: Exploratory
Object: aov
Parameter: group means
Method: generalized adjusted fractional Bayes factor

Posterior probabilities:
                            Pr(=0) Pr(<0) Pr(>0)
(Intercept)                  0.808  0.128  0.064
anchorrounded                0.000  0.000  1.000
motivationlow                0.000  1.000  0.000
anchorrounded:motivationlow  0.144  0.851  0.005

Main effects:
           Pr(null) Pr(alt)
anchor            0       1
motivation        0       1

Interaction effects:
                  Pr(null) Pr(alt)
anchor:motivation    0.251   0.749
\end{verbatim}
The results show clear evidence that there there is a main effect for the \texttt{anchor} factor and a main effect for the \texttt{motivation} factor (with posterior probabilities of approximately 1). Furthermore, there is some evidence that there interaction effect between the two factors is present (with a posterior probability of 0.749). More data need to be collected in order to draw a more decisive conclusion regarding the existence of an interaction.

It is interesting to see how the posterior probabilities for the hypotheses from the exploratory test relate to the p values of the classical significance tests of $H_0:\beta=0$ versus $H_1:\beta\not =0$. This can be checked by running:
\begin{verbatim}
R> ct <- lmtest::coeftest(aov1)
\end{verbatim}
This results in p values of $0.66697$, $1.842$e$-10$, $1.410$e$-06$, $0.01115$ for \texttt{(Intercept)},\\ \texttt{anchorrounded}, \texttt{motivationlow}, and \texttt{anchorrounded:motivationlow}, respectively. Thus, when using a significance level of 0.05 we would conclude that there is enough evidence in the data to conclude that an interaction effect is present. The Bayesian test on the other hand suggests that we have to be more cautious as there is still a posterior probability of 0.144 that no interaction effect is present. This illustrates that Bayes factor tests are generally more conservative than classical significance tests, which is one of the reasons for the recent recommendations to use smaller significance levels than the traditional choice of 0.05 \citep{Benjamin:2018}.

\subsection{Application 3: Testing group variances in neuropsychology}
\citet{Silverstein1995} conducted a psychological study to compare the attentional performances of 17 Tourette's syndrome (TS) patients, 17 ADHD patients, and 17 control subjects who did not suffer from TS or ADHD. The participants were shown a total of 120 sequences of either 3 or 12 letters. Each sequence contained either the letter T or the letter F at a random position. Each sequence was presented for 55 milliseconds and afterwards the participants had to indicate as quickly as possible whether the shown sequence contained a T or an F. After a participant completed all 120 sequences, his or her accuracy was calculated as the percentage of correct answers. In this section, we are interested in comparing the variances of the accuracies in the three groups. Research has shown that ADHD patients tend to be more variable in their attentional performances than subjects who do not suffer from ADHD \citep[e.g.,][]{Kofler2013, Russell2006}.  It is less well documented whether TS patients are less or more variable in their attentional performances than healthy control subjects. We will therefore test the following set of hypotheses to investigate whether TS patients are as variable in their attentional performances as either ADHD patients or healthy controls (C): $H_1\colon \sigma_C^2 = \sigma_{TS}^2 < \sigma_{ADHD}^2$ and $H_2\colon \sigma_C^2 < \sigma_{TS}^2 = \sigma_{ADHD}^2$. We will test these hypotheses against the null hypothesis stating equality of variances, $H_0\colon \sigma_C^2 = \sigma_{TS}^2 = \sigma_{ADHD}^2$, as well as the complement of the three aforementioned hypotheses given by $H_3\colon \neg \, (H_0 \lor H_1 \lor H_2)$. We include the complement to safeguard against the data supporting neither of $(H_0, H_1, H_2)$.

\citet{Silverstein1995} reported the following sample variances of the accuracies in the three groups: $s_C^2 = 15.52$, $s_{TS}^2 = 20.07$, and $s_{ADHD}^2 = 38.81$. The data are contained in a dataset called \verb|attention|. In \textbf{BFpack}, we can conduct the multiple hypothesis test and weigh the evidence in favor of the four hypotheses as follows:
\begin{verbatim}
R> bartlett <- bartlett_test(x = attention$accuracy, g = attention$group)
R> hypothesis <- c("Controls = TS < ADHD; Controls < TS = ADHD;
+    Controls = TS = ADHD")
R> set.seed(358)
R> BF_var <- BF(bartlett, hypothesis)
\end{verbatim}
Note that we use equal prior probabilities of the hypotheses by omitting the \texttt{prior} argument in the call to the \texttt{BF} function. The exploratory posterior probabilities for homogeneity of group variances can be obtained by running \texttt{summary(BF\_var)} which yields
\begin{verbatim}
Bayesian hypothesis test
Type: Exploratory
Object: BF_bartlett
Parameter: group variances
Method: generalized adjusted fractional Bayes factor

   homogeneity of variances no homogeneity of variances 
                      0.803                       0.197
\end{verbatim}
This results in evidence for equality of group variances. Note that the $p$ value in the classical Bartlett test for these data equals 0.1638 which implies that the null hypothesis of homogeneity of variances cannot be rejected using common significance levels, such as 0.05 or 0.01. Note however that the this $p$ value cannot be used as a measure for the evidence in the data in favor of homogeneity of group variances. This can be done using the proposed Bayes factor test which shows that the probability that the variances are equal is approximately 0.803.

The confirmatory test provides a more detailed analysis about the most plausible relationship between the hypotheses (also obtained using the \texttt{summary()} call):
\begin{verbatim}
Bayesian hypothesis test
Type: Confirmatory
Object: BF_bartlett
Parameter: group variances
Method: generalized adjusted fractional Bayes factor

Posterior probabilities:
   Pr(hypothesis|data)
H1               0.426
H2               0.278
H3               0.238
H4               0.058

Hypotheses:
H1: Controls=TS<ADHD
H2: Controls<TS=ADHD
H3: Controls=TS=ADHD
H4: complement
\end{verbatim}
Thus, $H_1$ receives strongest support from the data, but $H_2$ and $H_3$ are viable competitors. It appears that even the complement $H_3$ cannot be ruled out entirely given a posterior probability of 0.058. To conclude, the results indicate that TS population are as heterogeneous in their attentional performances as the healthy control population in this specific task, but further research would be required to obtain more conclusive evidence.

\subsection{Application 4: Multivariate linear regression in fMRI studies}
It is well established that the fusiform facial area (FFA), located in the inferior temporal cortex of the brain, plays an important role in the recognition of faces. This data comes from a study on the association between the thickness of specific cortical layers of the FFA and individual differences in the ability to recognize faces and vehicles \citep{McGuigin:2019}. High-resolution fMRI was recorded from 13 adult participants, after which the thickness of the superficial, middle, and deep layers of the FFA was quantified for each individual. In addition, individual differences in  face and vehicle recognition ability were assessed using a battery of tests.

\subsubsection{Analysis of the complete data}
In this example, two alternative hypotheses are tested.  In a recent study, \cite{McGuigin:2016} found that individual differences in the overall thickness of the FFA are negative correlated with the ability to recognize faces but positively correlated with the ability to recognize cars. $(H_{1})$ is the most parsimonious extension of these findings. It specifies that the magnitude and direction of the association between object recognition and layer thickness is not moderated by layer. To elaborate, consider a multivariate multiple regression model model with cortical thickness measures for the superficial, middle, and deep layers as three repeated (dependent) measures for each participant 
, and facial recognition ability and vehicle recognition ability as two dependent variables. Hypothesis $H_{1}$ is a main effects only model specifying that only main effect terms for face and vehicle are sufficient to predict the thickness of layers. The absence of layer $\times$ face or layer $\times$ vehicle interaction terms means that the relations between face and vehicle recognition are invariant across cortical layers. In other words, this hypothesis specifies that:
\begin{eqnarray*}
H_1&:&\beta_{Deep\_on\_Face} = \beta_{Middle\_on\_Face} = \beta_{Superficial\_on\_Face} < 0 <
\beta_{Deep\_on\_Vehicle}\\
&&= \beta_{Middle\_on\_Vehicle} = \beta_{Superficial\_on\_Vehicle}.
\end{eqnarray*}
That is, regression coefficients between face recognition and cortical thickness measures are expected to be negative, coefficients between vehicle recognition and cortical thickness measures are expected to be positive, and no layer-specific effect is expected for either faces or vehicles.

Hypothesis $H_{2}$ is based on prior findings concerning the early development of facial recognition abilities and the more rapid development of the deep layer of the FFA. This evidence leads to the following hypothesis: 
\begin{eqnarray*}
H_2&:&\beta_{Deep\_on\_Face} < \beta_{Middle\_on\_Face} =\beta_{Superficial\_on\_Face} < 0 < \beta_{Deep\_on\_Vehicle} \\
&&=\beta_{Middle\_on\_Vehicle} =\beta_{Superficial\_on\_Vehicle} 
\end{eqnarray*}
That is, the negative effect between facial recognition and the cortical thickness would be more pronounced in the deep layer, relative to the superficial and middle layers. One could attempt to test and compare these two hypotheses using linear mixed effects models software (e.g., the \texttt{gls} function in the \textbf{lme} package in \textsf{R}) with an appropriate covariance structure on the residuals to account for within-subject dependence. Alternatively one could use a model selection framework like that embodied in the \textbf{BayesFactor} package in \textsf{R}. Unfortunately, while these approaches can test some components of each hypothesis, they are not well suited to test the directional component of $H_1$, which specifies that all coefficients involving faces are smaller than 0 and that all coefficients involving vehicles are larger than 0. This hypothesis can, however, be tested using \textbf{BFpack} in the following way:
\begin{verbatim}
R> fmri.lm <- lm(cbind(Superficial, Middle, Deep) ~ Face + Vehicle,
+    data = fmri)
R> constraints.fmri <- "Face_on_Deep = Face_on_Superficial = Face_on_Middle
+    < 0 < Vehicle_on_Deep = Vehicle_on_Superficial = Vehicle_on_Middle;
+    Face_on_Deep < Face_on_Superficial = Face_on_Middle < 0 <
+    Vehicle_on_Deep = Vehicle_on_Superficial = Vehicle_on_Middle"
R> set.seed(123)
R> BF_fmri <- BF(fmri.lm, hypothesis = constraints.fmri)
R> summary(BF_fmri)
\end{verbatim}
This results in the following posterior probabilities and evidence matrix:
\begin{verbatim}
Posterior probabilities:
   Pr(hypothesis|data)
H1               0.023
H2               0.975
H3               0.002

Evidence matrix:
       H1    H2     H3
H1  1.000 0.024  13.35
H2 42.391 1.000 565.93
H3  0.075 0.002   1.00
\end{verbatim}
In this analysis, hypothesis $H_3$ is the complement hypothesis. The evidence matrix reveals there is clear evidence for $H_{2}$ against $H_1$ ($B_{21} = 42.391$) and extreme evidence for $H_2$ against $H_3$ ($B_{23} = 565.93$). The same conclusion can be drawn when looking at the posterior probabilities for the hypotheses. Based on these result we would conclude that hypothesis $H_2$ receives most evidence and the Bayesian probability of drawing the wrong conclusion after observing the data would be relatively small, namely, 0.025.

\subsubsection{Analysis with missing observations}\label{SectionMissing}
Here we illustrate how a Bayes factor test can be executed in the case of missing observations in the fMRI data set that are missing at random. A slightly simpler hypothesis test is considered to reduce the computation time\footnote{The hypotheses from Section 4.4 \textit{Analysis of complete data} has constraints on the effects across different predictor variables and different dependent variables, therefore requiring Monte Carlo estimation to obtain the Bayes factors. On the other hand, when the constraints are formulated on the effects of the same predictor on different dependent variables, an analytic expression is available for the Bayes factors.}
\begin{verbatim}
R> constraints.fmri2 <- 
+    "Face_on_Deep = Face_on_Superficial = Face_on_Middle < 0;
+    Face_on_Deep < Face_on_Superficial = Face_on_Middle < 0"
\end{verbatim}
First the Bayes factors and posterior probabilities are obtained for this hypothesis test for the complete data set:
\begin{verbatim}
R> fmri.lm2 <- lm(cbind(Superficial,Middle,Deep) ~ Face +
+    Vehicle, data = fmri)
R> BF.fmri2 <- BF(fmri.lm2, hypothesis = constraints.fmri2)
\end{verbatim}
This results in posterior probabilities of 0.050, 0.927, and 0.023 for the two constrained hypotheses and the complement hypothesis, respectively. The Bayes factor of the most supported hypothesis ($H_2$) against the second most supported hypothesis ($H_1$) equals $B_{21}=18.443$.

Now 10 missing observations (out of 65 separate observations in total) are randomly created that are missing at random:
\begin{verbatim}
R> fmri_missing <- fmri
R> set.seed(1234)
R> for(i in 1:10){
+    fmri_missing[sample(1:nrow(fmri), 1), sample(1:ncol(fmri), 1)] <- NA
+  }
\end{verbatim}
This results in 7 rows with at least one missing observation. Therefore listwise deletion would leave us with only 6 complete observations (of the 13 rows in total).
Even though list-wise deletion is generally not recommended \citep{Rubin:1987,Rubin:1996}, for this illustration we compute the Bayes factors and posterior probabilities based on these 6 complete data observations to illustrate the loss of evidence as a result of list-wise deletion.
\begin{verbatim}
R> fmri_listdel <- fmri_missing[!is.na(apply(fmri_missing, 1, sum)),]
R> fmri.lm2_listdel <- lm(cbind(Superficial, Middle, Deep) ~ Face + Vehicle,
+    data = fmri_listdel)
R> BF.fmri2_listdel <- BF(fmri.lm2_listdel, hypothesis = constraints.fmri2)
R> print(BF.fmri2_listdel)
\end{verbatim}
This results in posterior probabilities of 0.010, 0.820, and 0.170 for the two constrained hypotheses and the complement hypothesis, respectively. As expected the evidence for the hypothesis $H_2$ which received most evidence in based on the complete data set, decreased from 0.927 to 0.820.

Next we illustrate that multiple imputation results a smaller loss in evidence because the partly observed cases are still used in the analysis. We first generate 500 imputed data sets using \texttt{mice} from the \textbf{mice} package \citep{mice:2019}, and then use \texttt{BF()} to get the measures of relative fit and relative complexity for the equality and order constraints for the three hypotheses. These are be obtained from the element \texttt{BFtable\_confirmatory} of an object of class \texttt{BF}\footnote{Note that the measures of relative fit and relative complexity can also be found in the `Specification table' when calling the \texttt{summary()} function on an object of class \texttt{BF} in the case of a confirmatory test on the hypotheses specified in the \texttt{hypothesis} argument of the \texttt{BF()} function.}
\begin{verbatim}
R> M <- 500
R> library(mice)
R> mice_fmri <- mice :: mice(data = fmri_missing, m = M, meth = c("norm",
+    "norm", "norm", "norm", "norm"), diagnostics = F, printFlag = F)
R> relmeas_all <- matrix(unlist(lapply(1:M, function(m){
+    fmri.lm_m <- lm(cbind(Superficial, Middle, Deep) ~ Face + Vehicle,
+    data = mice::complete(mice_fmri, m))
+    BF.fmri2_m <- BF(fmri.lm_m, hypothesis = constraints.fmri2)
+    c(BF.fmri2_m$BFtable_confirmatory[, 1:4])
+  })),ncol = M)
R> relmeas <- matrix(apply(relmeas_all, 1, mean),nrow = 3)
R> row.names(relmeas) <- c("H1", "H2", "H3")
R> colnames(relmeas) <- c("comp_E", "comp_O", "fit_E", "fit_O")
R> BF_tu_confirmatory <- relmeas[,3] * relmeas[,4] / (relmeas[,1] *
+    relmeas[,2])
R> PHP <- BF_tu_confirmatory / sum(BF_tu_confirmatory)
R> print(PHP)
\end{verbatim}
This results in posterior probabilities of 0.066, 0.909, and 0.025 for the two constrained hypotheses and the complement hypothesis, respectively. Thus the posterior probability for the most supported hypothesis is still 0.909 (compared to 0.927 based on the complete data set), which is considerably larger than the posterior probability of 0.820 which was obtained using the data set after list-wise deletion.

\subsection{Application 5: Logistic regression in forensic psychology}\label{Application5}
The presence of systematic biases in the legal system runs counter to society's expectation of fairness. Moreover such biases can have profound personal ramifications, and the topic therefore warrants close scrutiny. \cite{Wilson:2015} examined the correlation between perceived facial trustworthiness and criminal-sentencing outcomes (data available at \url{https://osf.io/7mazn/}). In Study 1 photos of inmates who had been sentenced to death (or not) were rated by different groups of participants on trustworthiness, `Afrocentricity' (how sterotypically `black' participants were perceived as), attractiveness and facial maturity. Each photo was also coded for the presence of glasses/tattoos and facial width-to-height ratio. A logistic regression with sentencing as outcome was fitted to the predictors.

Previous research had shown that the facial width-to-height ratio (fWHR) has a postive effect on perceived aggression and thus may also have a positive effect on sentencing outcomes. In addition, perceived Afrocentricity had been shown to be associated with harsher sentences \citep[][]{Wilson:2015}. 
In the first hypothesis it was expected that all three predictors have a positive effect on the probability of being sentenced to death. Additionally, we might expect lack of perceived trustworthiness to have the largest effect. In the second hypothesis it was assumed that only trustworthiness has a positive effect. Finally, the complement hypothesis was considered. The hypotheses can then be summarized as follows
\begin{eqnarray*}
H_1&:& \beta_{trust} > (\beta_{fWHR}, \beta_{afro}) > 0\\
H_2&:& \beta_{trust} > \beta_{fWHR} = \beta_{afro} = 0\\
H_3&:& \text{neither $H_1$, nor $H_2$.}
\end{eqnarray*}

Before fitting the logistic regression we reverse-coded the trustworthiness scale and standardized it to be able to compare the magnitude the three effects. We can then test these hypotheses using \textbf{BFpack} and the fitted \texttt{glm} object from \textsf{R}. Note that the fitted object also contains covariates. The full logistic regression model was first fitted, and then the above hypotheses were tested on the fitted \texttt{glm} object:
\begin{verbatim}
R> fit <- glm(sent ~ ztrust + zfWHR + zAfro + glasses + attract +
+    maturity + tattoos, family = binomial(), data = wilson)
R> set.seed(123)
R> BF_glm <- BF(fit, hypothesis = "ztrust > (zfWHR, zAfro) > 0;
+    ztrust > zfWHR = zAfro = 0")
R> summary(BF_glm)
\end{verbatim}

In the output we see little support for the first two hypotheses; the complement receives most support:
\begin{verbatim}
Hypotheses:
   Pr(hypothesis)
H1               0.078
H2               0.002
H3               0.920

Hypotheses:
H1: ztrust>(zfWHR,zAfro)>0
H2: ztrust>(zfWHR,zAfro)=0
H3: complement
\end{verbatim}
The evidence matrix shows that the complement hypothesis is around 11.755 times as likely as the second best hypothesis:
\begin{verbatim}
Evidence matrix:
       H1      H2    H3
H1  1.000  36.193 0.085
H2  0.027   1.000 0.002
H3 11.755 433.890 1.000
\end{verbatim}
Based on these results we see that the complement receives most evidence. The fact that none of the two anticipated hypotheses were supported by the data indicates that the theory is not yet well-developed. Closer inspection of the beta-coefficients reveals that this is largely driven by the negative effect between perceived Afrocentricity and sentencing harshness ($\hat{\beta}_{zAfro}=-0.18071$). This unexpected result is discussed further by \cite{Wilson:2015} in their Supplementary Materials (\url{https://journals.sagepub.com/doi/suppl/10.1177/0956797615590992}).

Finally we illustrate how an exploratory Bayes factor test can be executed from output of a classical significance test:
\begin{verbatim}
R> ct <- lmtest::coeftest(fit)
R> BF(ct[,1], Sigma = diag(ct[,2]^2), n = nrow(wilson))
\end{verbatim}
Note that the dependency between the parameters can be ignored when performing tests on separate coefficients. The output is given by:
\begin{verbatim}
Call:
BF(ct[, 1], Sigma = diag(ct[, 2]^2), n = nrow(wilson))

Bayesian hypothesis test
Type: Exploratory
Object: numeric
Parameter: General
Method: Bayes factor using Gaussian approximations

Posterior probabilities
            Pr(=0) Pr(<0) Pr(>0)
(Intercept) 0.853  0.014  0.133
ztrust      0.000  0.000  1.000
zfWHR       0.001  0.000  0.999
zAfro       0.365  0.631  0.004
glasses     0.712  0.009  0.278
attract     0.930  0.041  0.029
maturity    0.770  0.219  0.011
tattoos     0.787  0.011  0.202
\end{verbatim}

\subsection{Application 6: Testing measures of association in neuropsychology}
Schizophrenia is often conceptualized as a disorder of ``dysconnectivity'' characterized by disruption in neural circuits that connect different regions of the brain (e.g., Friston \& Firth, 1995). This data set (originally collected by Ichinose, Han, Polyn, Park  and Tomarken (2019; summarized in Tomarken \& Mulder, in preparation) can be used to test whether such dysconnection is manifested behaviorally as weaker correlations among measures that we would expect to be highly correlated among non-schizophrenic individuals. 20 patients suffering from schizophrenia (SZ group) and 20 healthy control (HC group) participants were administered six measures of working memory. Ichinose et al. hypothesized that each of the 15 correlations would be smaller in the schizophrenic group relative to the control group.

This data set is an interesting case of how an order-constrained Bayesian approach can provide a more powerful and more appropriate test relative to alternative methods. Table \ref{tablecorr} presents the Pearson correlations for the two groups. Several features are notable: (1) Each of the 15 correlations is higher in the HC group than the SZ group; (2) On average the correlations among the HC group are rather high (on average $0.59$); and, (3) The average correlation within the SZ group is essentially 0. Despite this clear pattern, there were significant differences between the HC and SZ groups on only 2 of 15 correlations when the false discovery rate was used to control for multiple testing.

\begin{table}[ht]
\caption{Correlations for the SZ (above diagonal) and HC (below diagonal) groups.}
\centering
\newcommand{\mysubscript}[1]{\raisebox{-0.34ex}{\scriptsize#1}}
\renewcommand\thetable{3}
\begin{tabular}{rrrrrrr}
  \hline
 & Im & Del & Wmn & Cat & Fas & Rat \\ 
  \hline
  Im &  & 0.35 & -0.07 & -0.28 & -0.17 & 0.08 \\ 
  Del & 0.83 & & -0.22 & 0.16 & 0.27 & 0.09 \\ 
  Wmn & 0.65 & 0.50 &  & -0.05 & 0.01 & -0.02 \\ 
  Cat & 0.56 & 0.39 & 0.77 & & 0.22 & -0.25 \\ 
  Fas & 0.39 & 0.32 & 0.70 & 0.73  & &-0.14 \\ 
  Rat & 0.54 & 0.47 & 0.61 & 0.77 & 0.67 & \\ 
   \hline
\end{tabular}\label{tablecorr}
\end{table}

Given that the overall pattern of group differences is consistent with hypotheses, simultaneous testing procedures would appear to be a better approach than tests on individual correlations. Indeed, both maximum likelihood and resampling tests convincingly indicated that the covariance and correlation matrices across groups differ ($p < 0.01$). However, there are a number of ways in which two correlation or covariance matrices may differ. Thus, the conventional procedures for comparing matrices do not test the specific hypothesis that, for each of the 15 correlations, the value for the HC group is greater than the value for the SZ group.

This hypothesis can, however, be tested in a straightforward manner using \textbf{BFpack}. $H_{1}$ specifies that each correlation in the HC group is expected to be larger than the corresponding correlation in the SZ group (i.e., a total of 15 order constraints were imposed). $H_{A}$ represents any pattern of correlations other than those that were consistent with $H_1$. The R syntax is as follows:
\begin{verbatim}
R> lm6 <- lm(cbind(Im, Del, Wmn, Cat, Fas, Rat) ~ -1 + Group,
+    data = memory)
R> set.seed(123)
R> BF6_cor <- BF(lm6, hypothesis =
+    "Del_with_Im_in_GroupHC > Del_with_Im_in_GroupSZ &
+    Del_with_Wmn_in_GroupHC > Del_with_Wmn_in_GroupSZ & 
+    Del_with_Cat_in_GroupHC > Del_with_Cat_in_GroupSZ &
+    Del_with_Fas_in_GroupHC > Del_with_Fas_in_GroupSZ &
+    Del_with_Rat_in_GroupHC > Del_with_Rat_in_GroupSZ & 
+    Im_with_Wmn_in_GroupHC > Im_with_Wmn_in_GroupSZ & 
+    Im_with_Cat_in_GroupHC > Im_with_Cat_in_GroupSZ &
+    Im_with_Fas_in_GroupHC > Im_with_Fas_in_GroupSZ &
+    Im_with_Rat_in_GroupHC > Im_with_Rat_in_GroupSZ & 
+    Wmn_with_Cat_in_GroupHC > Wmn_with_Cat_in_GroupSZ &
+    Wmn_with_Fas_in_GroupHC > Wmn_with_Fas_in_GroupSZ &
+    Wmn_with_Rat_in_GroupHC > Wmn_with_Rat_in_GroupSZ & 
+    Cat_with_Fas_in_GroupHC > Cat_with_Fas_in_GroupSZ &
+    Cat_with_Rat_in_GroupHC > Cat_with_Rat_in_GroupSZ & 
+    Fas_with_Rat_in_GroupHC > Fas_with_Rat_in_GroupSZ")
R> summary(BF6_cor)
\end{verbatim}
Based on the summary, which can be obtained by running
the Bayes Factor for $H_{1}$ against $H_{A}$ was approximately $4631.01$ and the posterior probability for $H_{1}$ was effectively 1. Thus the order-constrained analysis indicate decisive support for the researchers' hypothesis.

\subsection{Application 7: Testing intraclass correlations in educational testing}
Data from the Trends in International Mathematics and Science Study (TIMSS; \url{http://www.iea.nl/timss}) were used to examine differences in intraclass correlations of four countries (The Netherlands (NL), Croatia (HR), Germany (DE), and Denmark (DK)) with respect to the mathematics achievements of fourth graders (e.g., the first plausible value was used as a measure of mathematics achievement). The sample design of the TIMSS data set is known to describe three levels with students nested within classrooms/schools, and classrooms/schools nested within countries (e.g., one classroom is sampled per school). In this example, the TIMSS 2011 assessment was considered.

The intraclass correlation was defined as the correlation among measured mathematics achievements of grade-4 students attending the same school. This intraclass correlation was assumed to be homogenous across schools in the same country, but was allowed to be different across countries. For the four countries, differences in intraclass correlations were tested using the Bayes factor. The size of the intraclass correlation can be of specific interest, since sampling becomes less efficient when the intraclass correlation increases. Countries with low intraclass correlations have fewer restrictions on the sample design, where countries with high intraclass correlations require more efficient sample designs, larger sample sizes, or both. Knowledge about the size of the heterogeneity provide useful information to optimize the development of a suitable sample design and to minimize the effects of high intraclass correlations.

The TIMSS data sample in \textbf{BFpack} consists of four countries, where data was retrieved from The Netherlands ($93,112$), Croatia ($139, 106$), Germany ($179, 170$), and Denmark ($166, 153$) with the sampled number of schools in brackets for 2011 and 2015, respectively. Differences in intraclass correlations were tested conditional on several student variables (e.g., gender, student sampling weight variable). The following hypotheses on intraclass correlations were considered in the analyses. Country-ordered intraclass correlations were considered by hypothesis $H_1$, equal (invariant) intra-class correlations were represented by hypothesis $H_2$, and their complement was specified as hypothesis $H_3$:
\begin{eqnarray*}
H_1 &:& \rho_{NL} < \rho_{HR} < \rho_{DE} < \rho_{DK} \\
H_2 &:& \rho_{NL} = \rho_{HR} = \rho_{DE} = \rho_{DK} \\
H_3 &:& \text{neither $H_1$, nor $H_2$.} 
\end{eqnarray*}
The ordering in the intraclass correlations was hypothesized by considering the reported standard errors of the country-mean scores. From the variance inflation factor followed, $1+(p-1)\rho$, with $p$ the number of students in each school (balanced design), it follows that the variance of the mean increases for increasing values of the intraclass correlation coefficient. As a result, the ordering in estimated standard errors of the average mathematic achievements of fourth graders of the cycles from 2003 to 2015 was used to hypothesis the order in intraclass correlations. From a more substantive perspective, it is expected that schools in the Netherlands do not differ much with respect to their performances (low intraclass correlation) in contrast to Denmark, where school performances may differ considerably (high intraclass correlation).

A linear mixed effects model was used to obtain (restricted) maximum likelihood estimates of the fixed effects of the student variables and the country means, the four random effects corresponding to the clustering of students in schools in each country, and the measurement error variance, given the 2011 assessment data.
\begin{verbatim}
R> library(lme4)
R> timssICC_subset <- timssICC[(timssICC$groupNL11 == 1) + 
+    (timssICC$groupHR11 == 1) + (timssICC$groupDE11 == 1) +
+    (timssICC$groupDK11 == 1) > 0,]
R> outlme1 <- lmer(math ~ -1 + gender + weight + lln +
+    groupNL11 + (0 + groupNL11 | schoolID) +
+    groupHR11 + (0 + groupHR11 | schoolID) +
+    groupDE11 + (0 + groupDE11 | schoolID) +
+    groupDK11 + (0 + groupDK11 | schoolID), 
+    data=timssICC_subset)
\end{verbatim}
where the \texttt{schoolID} factor variable assigns a unique code to each school, and each country-specific group variable (e.g., \texttt{groupNL11}) equals one when it concerns a school in that country and zero otherwise. The \texttt{lmer} output object\citep{lme4} was used as input in the BF function for the Bayes factor computation, where hypothesis $H_1$ and $H_2$ were added as arguments in the function call;
\begin{verbatim}
R> set.seed(123)
R> BFicc <- BF(outlme1, hypothesis = 
+    "groupNL11 < groupHR11 < groupDE11 < groupDK11;
+    groupNL11 = groupHR11 = groupDE11 = groupDK11")
\end{verbatim}
The output object contains the posterior mean and median estimates of the ICCs (obtained via \texttt{BFicc\$estimates}), which are represented in Table \ref{tableICC}. The REML intraclass correlation estimates are also given for each country, which followed directly from the random effect estimates of the \texttt{lmer} output. It can be seen that the posterior mean and REML estimates are quite close, and the REML estimates are also located between the 2.5\% and 97.5\% percentile estimates.

	\begin{table}[!t]
		\centering
		\caption{TIMSS 2011: Intraclass correlation estimates for NL, HR, DE, and DK}
		\label{tableICC}
		\begin{tabular}{ccccc}
			\hline
			\multicolumn{1}{c}{Statistic} & \multicolumn{1}{c}{NL} & \multicolumn{1}{c}{HR} & \multicolumn{1}{c}{DE} & \multicolumn{1}{c}{DK} \\ \hline
			REML     & 0.094 & 0.122 & 0.156 & 0.195 \\
            Mean    & 0.110  &   0.132   &  0.163     &0.201 \\
            Median & 0.107 &    0.131  &   0.161     &0.200 \\
            2.5\%   & 0.066 &    0.094  &   0.125  &   0.159 \\
            97.5\% & 0.172  &   0.181  &   0.207  &   0.251 \\ \hline
		\end{tabular}
	\end{table}

%
By running \texttt{summary(BFicc)} we get the results of the exploratory and confirmatory tests. 
The exploratory tests provide posterior probabilities of whether each intraclass correlation equals zero, negative, or positive. Evidence in favor of a negative intraclass correlation indicates that a multilevel model may not be appropriate for modeling these data \citep{MulderFox:2019}. As can be seen the exploratory results indicate that a multilevel model is a appropriate for these data:
\begin{verbatim}
Bayesian hypothesis test
Type: Exploratory
Object: lmerMod
Parameter: intraclass correlations
Method: Bayes factor based on uniform priors

          icc=0 icc<0 icc>0
groupNL11     0     0     1
groupHR11     0     0     1
groupDE11     0     0     1
groupDK11     0     0     1
\end{verbatim}
Furthermore the posterior probabilities of the specified hypotheses shows how our beliefs are updated in light of the observed data regarding the hypotheses that were formulated on the variation of school performance across countries.
\begin{verbatim}
Bayesian hypothesis test
Type: Confirmatory
Object: lmerMod
Parameter: intraclass correlations
Method: Bayes factor based on uniform priors

Posterior probabilities:
   Pr(hypothesis|data)
H1               0.509
H2               0.471
H3               0.020

Hypotheses:
H1: groupNL11<groupHR11<groupDE11<groupDK11
H2: groupNL11=groupHR11=groupDE11=groupDK11
H3: complement
\end{verbatim}
The posterior probabilities of the three hypotheses in the confirmatory test reveal that there is approximately equal plausibility for $H_2$ and $H_3$ to be true (with posterior probabilities of 0.509 and 0.471, respectively), and the complement hypothesis is unlike to be true (with a posterior probability of 0.020). It can be concluded that the data gave most support to an ordering of the intraclass correlations, where the Netherlands have the smallest intraclass correlation and Denmark the highest. The evidence however is practically equal to the evidence for the equality hypothesis. Efficient sampling strategies are needed in countries with positive intraclass correlations, where countries with higher intraclass correlations will benefit more from efficient stratification strategies.

\section{Concluding remarks}\label{concl}
The \textsf{R} package \textbf{BFpack} was designed to allow substantive researchers to perform Bayes factor tests via commonly used statistical functions in \textsf{R}, such as \texttt{lm}, \texttt{aov}, \texttt{hetcor}, or \texttt{glm}. Furthermore by specifying a simple string that captures the hypotheses of interest, users can make use of the flexibility of Bayes factors to simultaneously test multiple hypotheses which may involve equality as well as order constraints on the parameters of interest. This will allow users to move beyond traditional null hypothesis (significance) testing. In the near future the package will be extended by also including more complex statistical models such as structural equation models and generalized linear mixed models.


\section*{Acknowledgments}

The first author is supported by a Vidi grant from the Netherlands Organization of Scientific Research (NWO). Regarding the applications, Application 1 was provided by Xin Gu, Application 2 by Herbert Hoijtink, Application 3 by Florian B\"{o}ing-Messing, Application 4 by Andrew Tomarken, Application 5 by Anton Ollson Collentine, Application 6 by Andrew Tomarken, Application 7 by Jean-Paul Fox, and Application 8 by Marlyne Meijerink.

\bibliographystyle{apacite}
\bibliography{refs_mulder_copy.bib}

\begin{thebibliography}{}

\bibitem [\protect \citeauthoryear {%
Bartlett%
}{%
Bartlett%
}{%
{\protect \APACyear {1957}}%
}]{%
Bartlett:1957}
\APACinsertmetastar {%
Bartlett:1957}%
\begin{APACrefauthors}%
Bartlett, M.%
\end{APACrefauthors}%
\unskip\
\newblock
\APACrefYearMonthDay{1957}{}{}.
\newblock
{\BBOQ}\APACrefatitle {A {C}omment on {D}. {V}. {L}indley's {S}tatistical
  {P}aradox} {A {C}omment on {D}. {V}. {L}indley's {S}tatistical
  {P}aradox}.{\BBCQ}
\newblock
\APACjournalVolNumPages{Biometrika}{44}{}{533--534}.
\PrintBackRefs{\CurrentBib}

\bibitem [\protect \citeauthoryear {%
Bates%
\ \protect \BOthers {.}}{%
Bates%
\ \protect \BOthers {.}}{%
{\protect \APACyear {2019}}%
}]{%
lme4}
\APACinsertmetastar {%
lme4}%
\begin{APACrefauthors}%
Bates, D.%
, Maechler, M.%
, Boler, B.%
, Walker, S.%
, Christensen, R\BPBI H\BPBI B.%
, Singmann, H.%
\BDBL {}Fox, J.%
\end{APACrefauthors}%
\unskip\
\newblock
\APACrefYearMonthDay{2019}{}{}.
\newblock
{\BBOQ}\APACrefatitle {\textbf{lme4}: Linear Mixed-Effects Models using `Eigen'
  and S4} {\textbf{lme4}: Linear mixed-effects models using `eigen' and
  s4}{\BBCQ}\ [\bibcomputersoftwaremanual].
\newblock
\begin{APACrefURL}
  \url{https://cran.r-project.org/web/packages/lme4/index.html}
  \end{APACrefURL}
\newblock
\APACrefnote{\textsf{R} package version 1.1-21}
\PrintBackRefs{\CurrentBib}

\bibitem [\protect \citeauthoryear {%
Benjamin%
\ \protect \BOthers {.}}{%
Benjamin%
\ \protect \BOthers {.}}{%
{\protect \APACyear {2018}}%
}]{%
Benjamin:2018}
\APACinsertmetastar {%
Benjamin:2018}%
\begin{APACrefauthors}%
Benjamin, D\BPBI J.%
, Berger, J\BPBI O.%
, Johannesson, M.%
, Nosek, B\BPBI A.%
, Wagenmakers, E\BHBI J.%
, Berk, R.%
\BCBL {}\ \BBA {} {... Johnson}, V\BPBI E.%
\end{APACrefauthors}%
\unskip\
\newblock
\APACrefYearMonthDay{2018}{}{}.
\newblock
{\BBOQ}\APACrefatitle {Redefine Statistical Significance} {Redefine statistical
  significance}.{\BBCQ}
\newblock
\APACjournalVolNumPages{Nature Human Behaviour}{2}{}{6--10}.
\newblock
\begin{APACrefURL}
  \url{https://www.nature.com/articles/s41562%20017%200189%20z}
  \end{APACrefURL}
\PrintBackRefs{\CurrentBib}

\bibitem [\protect \citeauthoryear {%
Berger%
}{%
Berger%
}{%
{\protect \APACyear {2006}}%
}]{%
Berger:2006}
\APACinsertmetastar {%
Berger:2006}%
\begin{APACrefauthors}%
Berger, J\BPBI O.%
\end{APACrefauthors}%
\unskip\
\newblock
\APACrefYearMonthDay{2006}{}{}.
\newblock
{\BBOQ}\APACrefatitle {The {C}ase for {O}bjective {B}ayesian {A}nalysis} {The
  {C}ase for {O}bjective {B}ayesian {A}nalysis}.{\BBCQ}
\newblock
\APACjournalVolNumPages{Bayesian Analysis}{1}{}{385--402}.
\newblock
\begin{APACrefDOI} \doi{10.1214/06-BA115} \end{APACrefDOI}
\PrintBackRefs{\CurrentBib}

\bibitem [\protect \citeauthoryear {%
Berger%
\ \BBA {} Delampady%
}{%
Berger%
\ \BBA {} Delampady%
}{%
{\protect \APACyear {1987}}%
}]{%
Berger:1987}
\APACinsertmetastar {%
Berger:1987}%
\begin{APACrefauthors}%
Berger, J\BPBI O.%
\BCBT {}\ \BBA {} Delampady, M.%
\end{APACrefauthors}%
\unskip\
\newblock
\APACrefYearMonthDay{1987}{}{}.
\newblock
{\BBOQ}\APACrefatitle {Testing {P}recise {H}ypotheses} {Testing {P}recise
  {H}ypotheses}.{\BBCQ}
\newblock
\APACjournalVolNumPages{Statistical Science}{2}{}{317--335}.
\newblock
\begin{APACrefDOI} \doi{10.1214/ss/1177013238} \end{APACrefDOI}
\PrintBackRefs{\CurrentBib}

\bibitem [\protect \citeauthoryear {%
Berger%
\ \BBA {} Pericchi%
}{%
Berger%
\ \BBA {} Pericchi%
}{%
{\protect \APACyear {2001}}%
}]{%
Berger:2001}
\APACinsertmetastar {%
Berger:2001}%
\begin{APACrefauthors}%
Berger, J\BPBI O.%
\BCBT {}\ \BBA {} Pericchi, L.%
\end{APACrefauthors}%
\unskip\
\newblock
\APACrefYearMonthDay{2001}{}{}.
\newblock
{\BBOQ}\APACrefatitle {Objective {B}ayesian {M}ethods for {M}odel {S}election:
  {I}ntroduction and {C}omparison (with {D}iscussion)} {Objective {B}ayesian
  {M}ethods for {M}odel {S}election: {I}ntroduction and {C}omparison (with
  {D}iscussion)}.{\BBCQ}
\newblock
\BIn{} P.~Lahiri\ (\BED), \APACrefbtitle {Model Selection} {Model selection}\
  (\BPGS\ 135--207).
\newblock
\APACaddressPublisher{}{Hayward, CA: Institute of Mathematical Statistics}.
\PrintBackRefs{\CurrentBib}

\bibitem [\protect \citeauthoryear {%
{B\"{o}ing-Messing}%
\ \BBA {} Mulder%
}{%
{B\"{o}ing-Messing}%
\ \BBA {} Mulder%
}{%
{\protect \APACyear {2016}}%
}]{%
BoeingMessing:2016}
\APACinsertmetastar {%
BoeingMessing:2016}%
\begin{APACrefauthors}%
{B\"{o}ing-Messing}, F.%
\BCBT {}\ \BBA {} Mulder, J.%
\end{APACrefauthors}%
\unskip\
\newblock
\APACrefYearMonthDay{2016}{}{}.
\newblock
{\BBOQ}\APACrefatitle {Automatic Bayes Factors for Testing Variances of Two
  Independent normal distributions} {Automatic bayes factors for testing
  variances of two independent normal distributions}.{\BBCQ}
\newblock
\APACjournalVolNumPages{Journal of Mathematical Psychology}{72}{}{158--170}.
\newblock
\begin{APACrefDOI} \doi{10.1016/j.jmp.2015.08.001} \end{APACrefDOI}
\PrintBackRefs{\CurrentBib}

\bibitem [\protect \citeauthoryear {%
{B\"{o}ing-Messing}%
\ \BBA {} Mulder%
}{%
{B\"{o}ing-Messing}%
\ \BBA {} Mulder%
}{%
{\protect \APACyear {2017}}%
}]{%
BoingMessing:2017}
\APACinsertmetastar {%
BoingMessing:2017}%
\begin{APACrefauthors}%
{B\"{o}ing-Messing}, F.%
\BCBT {}\ \BBA {} Mulder, J.%
\end{APACrefauthors}%
\unskip\
\newblock
\APACrefYearMonthDay{2017}{}{}.
\newblock
{\BBOQ}\APACrefatitle {Bayesian Evaluation of Constrained Hypotheses on
  Variances of Multiple Independent Groups} {Bayesian evaluation of constrained
  hypotheses on variances of multiple independent groups}.{\BBCQ}
\newblock
\APACjournalVolNumPages{Psychological Methods}{22}{2}{262--287}.
\newblock
\begin{APACrefDOI} \doi{10.1037/met0000116} \end{APACrefDOI}
\PrintBackRefs{\CurrentBib}

\bibitem [\protect \citeauthoryear {%
{B\"{o}ing-Messing}%
\ \BBA {} Mulder%
}{%
{B\"{o}ing-Messing}%
\ \BBA {} Mulder%
}{%
{\protect \APACyear {2018}}%
}]{%
BoeingMessing:2018}
\APACinsertmetastar {%
BoeingMessing:2018}%
\begin{APACrefauthors}%
{B\"{o}ing-Messing}, F.%
\BCBT {}\ \BBA {} Mulder, J.%
\end{APACrefauthors}%
\unskip\
\newblock
\APACrefYearMonthDay{2018}{}{}.
\newblock
{\BBOQ}\APACrefatitle {Automatic {B}ayes Factors for Testing Equality- and
  Inequality-Constrained Hypotheses on Variances} {Automatic {B}ayes factors
  for testing equality- and inequality-constrained hypotheses on
  variances}.{\BBCQ}
\newblock
\APACjournalVolNumPages{Psychometrika}{83}{3}{586--617}.
\newblock
\begin{APACrefDOI} \doi{10.1007/s11336-018-9615-z} \end{APACrefDOI}
\PrintBackRefs{\CurrentBib}

\bibitem [\protect \citeauthoryear {%
Box%
\ \BBA {} Tiao%
}{%
Box%
\ \BBA {} Tiao%
}{%
{\protect \APACyear {1973}}%
}]{%
BoxTiao}
\APACinsertmetastar {%
BoxTiao}%
\begin{APACrefauthors}%
Box, G\BPBI E\BPBI P.%
\BCBT {}\ \BBA {} Tiao, G\BPBI C.%
\end{APACrefauthors}%
\unskip\
\newblock
\APACrefYear{1973}.
\newblock
\APACrefbtitle {Bayesian Inference in Statistical Snalysis} {Bayesian inference
  in statistical snalysis}.
\newblock
\APACaddressPublisher{}{Reading, MA: Addison-Wesley}.
\PrintBackRefs{\CurrentBib}

\bibitem [\protect \citeauthoryear {%
Braeken%
, Mulder%
\BCBL {}\ \BBA {} Wood%
}{%
Braeken%
\ \protect \BOthers {.}}{%
{\protect \APACyear {2015}}%
}]{%
Braeken:2015}
\APACinsertmetastar {%
Braeken:2015}%
\begin{APACrefauthors}%
Braeken, J.%
, Mulder, J.%
\BCBL {}\ \BBA {} Wood, S.%
\end{APACrefauthors}%
\unskip\
\newblock
\APACrefYearMonthDay{2015}{}{}.
\newblock
{\BBOQ}\APACrefatitle {Relative Effects at Work: Bayes Factors for Order
  Hypotheses} {Relative effects at work: Bayes factors for order
  hypotheses}.{\BBCQ}
\newblock
\APACjournalVolNumPages{Journal of Management}{41}{2}{544-573}.
\newblock
\begin{APACrefDOI} \doi{10.1177/0149206314525206} \end{APACrefDOI}
\PrintBackRefs{\CurrentBib}

\bibitem [\protect \citeauthoryear {%
Consonni%
\ \BBA {} Veronese%
}{%
Consonni%
\ \BBA {} Veronese%
}{%
{\protect \APACyear {2008}}%
}]{%
Consonni:2008}
\APACinsertmetastar {%
Consonni:2008}%
\begin{APACrefauthors}%
Consonni, G.%
\BCBT {}\ \BBA {} Veronese, P.%
\end{APACrefauthors}%
\unskip\
\newblock
\APACrefYearMonthDay{2008}{}{}.
\newblock
{\BBOQ}\APACrefatitle {Compatibility of Prior Distribution Across Linear
  Models} {Compatibility of prior distribution across linear models}.{\BBCQ}
\newblock
\APACjournalVolNumPages{Statistical Science}{23}{3}{332--353}.
\newblock
\begin{APACrefDOI} \doi{10.1214/08-STS258} \end{APACrefDOI}
\PrintBackRefs{\CurrentBib}

\bibitem [\protect \citeauthoryear {%
{De Santis}%
\ \BBA {} Spezzaferri%
}{%
{De Santis}%
\ \BBA {} Spezzaferri%
}{%
{\protect \APACyear {1999}}%
}]{%
DeSantis:1999}
\APACinsertmetastar {%
DeSantis:1999}%
\begin{APACrefauthors}%
{De Santis}, F.%
\BCBT {}\ \BBA {} Spezzaferri, F.%
\end{APACrefauthors}%
\unskip\
\newblock
\APACrefYearMonthDay{1999}{}{}.
\newblock
{\BBOQ}\APACrefatitle {Methods for Default and Robust {B}ayesian Model
  Comparison: the Fractional {B}ayes Factor Approach} {Methods for default and
  robust {B}ayesian model comparison: the fractional {B}ayes factor
  approach}.{\BBCQ}
\newblock
\APACjournalVolNumPages{International Statistical Review}{67}{3}{267-286}.
\newblock
\begin{APACrefDOI} \doi{10.1111/j.1751-5823.1999.tb00449.x} \end{APACrefDOI}
\PrintBackRefs{\CurrentBib}

\bibitem [\protect \citeauthoryear {%
{De Santis}%
\ \BBA {} Spezzaferri%
}{%
{De Santis}%
\ \BBA {} Spezzaferri%
}{%
{\protect \APACyear {2001}}%
}]{%
DeSantis:2001}
\APACinsertmetastar {%
DeSantis:2001}%
\begin{APACrefauthors}%
{De Santis}, F.%
\BCBT {}\ \BBA {} Spezzaferri, F.%
\end{APACrefauthors}%
\unskip\
\newblock
\APACrefYearMonthDay{2001}{}{}.
\newblock
{\BBOQ}\APACrefatitle {Consistent Fractional Bayes Factors for Nested Normal
  Linear Models} {Consistent fractional bayes factors for nested normal linear
  models}.{\BBCQ}
\newblock
\APACjournalVolNumPages{Journal of Statistical Planning and
  Inference}{97}{2}{305-321}.
\newblock
\begin{APACrefDOI} \doi{10.1016/S0378-3758(00)00240-8} \end{APACrefDOI}
\PrintBackRefs{\CurrentBib}

\bibitem [\protect \citeauthoryear {%
de Jong%
, Rigotti%
\BCBL {}\ \BBA {} Mulder%
}{%
de Jong%
\ \protect \BOthers {.}}{%
{\protect \APACyear {2017}}%
}]{%
DeJong:2017}
\APACinsertmetastar {%
DeJong:2017}%
\begin{APACrefauthors}%
de Jong, J.%
, Rigotti, T.%
\BCBL {}\ \BBA {} Mulder, J.%
\end{APACrefauthors}%
\unskip\
\newblock
\APACrefYearMonthDay{2017}{}{}.
\newblock
{\BBOQ}\APACrefatitle {One After the Other: Effects of Sequence Patterns of
  Breached and Overfulfilled Obligations} {One after the other: Effects of
  sequence patterns of breached and overfulfilled obligations}.{\BBCQ}
\newblock
\APACjournalVolNumPages{European Journal of Work and Organizational
  Psychology}{26}{3}{337--355}.
\newblock
\begin{APACrefDOI} \doi{10.1080/1359432X.2017.1287074} \end{APACrefDOI}
\PrintBackRefs{\CurrentBib}

\bibitem [\protect \citeauthoryear {%
Dickey%
}{%
Dickey%
}{%
{\protect \APACyear {1971}}%
}]{%
Dickey:1971}
\APACinsertmetastar {%
Dickey:1971}%
\begin{APACrefauthors}%
Dickey, J.%
\end{APACrefauthors}%
\unskip\
\newblock
\APACrefYearMonthDay{1971}{}{}.
\newblock
{\BBOQ}\APACrefatitle {The Weighted Likelihood Ratio, Linear Hypotheses on
  Normal Location Parameters} {The weighted likelihood ratio, linear hypotheses
  on normal location parameters}.{\BBCQ}
\newblock
\APACjournalVolNumPages{The Annals of Statistics}{42}{1}{204--223}.
\PrintBackRefs{\CurrentBib}

\bibitem [\protect \citeauthoryear {%
Dogge%
, Gayet%
, Custers%
, Hoijtink%
\BCBL {}\ \BBA {} Aarts%
}{%
Dogge%
\ \protect \BOthers {.}}{%
{\protect \APACyear {2019}}%
}]{%
Digge:2019}
\APACinsertmetastar {%
Digge:2019}%
\begin{APACrefauthors}%
Dogge, M.%
, Gayet, S.%
, Custers, R.%
, Hoijtink, H.%
\BCBL {}\ \BBA {} Aarts, H.%
\end{APACrefauthors}%
\unskip\
\newblock
\APACrefYearMonthDay{2019}{}{}.
\newblock
{\BBOQ}\APACrefatitle {Perception of Action-Outcomes is Shaped by Life-Long and
  Contextual Expectations} {Perception of action-outcomes is shaped by
  life-long and contextual expectations}.{\BBCQ}
\newblock
\APACjournalVolNumPages{Scientific Reports}{9}{}{1--9}.
\PrintBackRefs{\CurrentBib}

\bibitem [\protect \citeauthoryear {%
Flore%
, Mulder%
\BCBL {}\ \BBA {} Wicherts%
}{%
Flore%
\ \protect \BOthers {.}}{%
{\protect \APACyear {2019}}%
}]{%
Flore:2019}
\APACinsertmetastar {%
Flore:2019}%
\begin{APACrefauthors}%
Flore, P\BPBI C.%
, Mulder, J.%
\BCBL {}\ \BBA {} Wicherts, J\BPBI M.%
\end{APACrefauthors}%
\unskip\
\newblock
\APACrefYearMonthDay{2019}{}{}.
\newblock
{\BBOQ}\APACrefatitle {The Influence of Gender Stereotype Threat on Mathematics
  Test Scores of Dutch High School Students: A Registered Report} {The
  influence of gender stereotype threat on mathematics test scores of dutch
  high school students: A registered report}.{\BBCQ}
\newblock
\APACjournalVolNumPages{Comprehensive Results in Social Psychology}{}{}{}.
\newblock
\begin{APACrefDOI} \doi{10.1080/23743603.2018.1559647} \end{APACrefDOI}
\PrintBackRefs{\CurrentBib}

\bibitem [\protect \citeauthoryear {%
J.~Fox%
}{%
J.~Fox%
}{%
{\protect \APACyear {2016}}%
}]{%
polycor:2016}
\APACinsertmetastar {%
polycor:2016}%
\begin{APACrefauthors}%
Fox, J.%
\end{APACrefauthors}%
\unskip\
\newblock
\APACrefYearMonthDay{2016}{}{}.
\newblock
{\BBOQ}\APACrefatitle {\textbf{polycor}: Polychoric and Polyserial
  Correlations} {\textbf{polycor}: Polychoric and polyserial
  correlations}{\BBCQ}\ [\bibcomputersoftwaremanual].
\newblock
\begin{APACrefURL}
  \url{https://cran.r-project.org/web/packages/polycor/index.html}
  \end{APACrefURL}
\newblock
\APACrefnote{\textsf{R} package version 0.7-9}
\PrintBackRefs{\CurrentBib}

\bibitem [\protect \citeauthoryear {%
J.~Fox%
\ \BBA {} Weisberg%
}{%
J.~Fox%
\ \BBA {} Weisberg%
}{%
{\protect \APACyear {2019}}%
}]{%
Fox:2019}
\APACinsertmetastar {%
Fox:2019}%
\begin{APACrefauthors}%
Fox, J.%
\BCBT {}\ \BBA {} Weisberg, S.%
\end{APACrefauthors}%
\unskip\
\newblock
\APACrefYearMonthDay{2019}{}{}.
\newblock
{\BBOQ}\APACrefatitle {\textbf{car}: Companion to Applied Regression}
  {\textbf{car}: Companion to applied regression}{\BBCQ}\
  [\bibcomputersoftwaremanual].
\newblock
\begin{APACrefURL} \url{https://cran.r-project.org/web/packages/car/index.html}
  \end{APACrefURL}
\newblock
\APACrefnote{\textsf{R} package version 3.0-3}
\PrintBackRefs{\CurrentBib}

\bibitem [\protect \citeauthoryear {%
J\BHBI P.~Fox%
, Mulder%
\BCBL {}\ \BBA {} Sinharay%
}{%
J\BHBI P.~Fox%
\ \protect \BOthers {.}}{%
{\protect \APACyear {2017}}%
}]{%
Fox:2017}
\APACinsertmetastar {%
Fox:2017}%
\begin{APACrefauthors}%
Fox, J\BHBI P.%
, Mulder, J.%
\BCBL {}\ \BBA {} Sinharay, S.%
\end{APACrefauthors}%
\unskip\
\newblock
\APACrefYearMonthDay{2017}{}{}.
\newblock
{\BBOQ}\APACrefatitle {Bayes Factor Covariance Testing in Item Response Models}
  {Bayes factor covariance testing in item response models}.{\BBCQ}
\newblock
\APACjournalVolNumPages{Psychometrika}{82}{4}{979--1006}.
\newblock
\begin{APACrefDOI} \doi{10.1007/s11336-017-9577-6} \end{APACrefDOI}
\PrintBackRefs{\CurrentBib}

\bibitem [\protect \citeauthoryear {%
Gancia-Donato%
\ \BBA {} Sun%
}{%
Gancia-Donato%
\ \BBA {} Sun%
}{%
{\protect \APACyear {2007}}%
}]{%
Garcia:2007}
\APACinsertmetastar {%
Garcia:2007}%
\begin{APACrefauthors}%
Gancia-Donato, G.%
\BCBT {}\ \BBA {} Sun, D.%
\end{APACrefauthors}%
\unskip\
\newblock
\APACrefYearMonthDay{2007}{}{}.
\newblock
{\BBOQ}\APACrefatitle {Objective Priors for Hypothesis Testing in One-Way
  Random Effects Models} {Objective priors for hypothesis testing in one-way
  random effects models}.{\BBCQ}
\newblock
\APACjournalVolNumPages{Canadian Journal of Statistics}{35}{}{302-320}.
\newblock
\begin{APACrefDOI} \doi{10.1002/cjs.5550350207} \end{APACrefDOI}
\PrintBackRefs{\CurrentBib}

\bibitem [\protect \citeauthoryear {%
Gelman%
\ \protect \BOthers {.}}{%
Gelman%
\ \protect \BOthers {.}}{%
{\protect \APACyear {2014}}%
}]{%
Gelman:2014}
\APACinsertmetastar {%
Gelman:2014}%
\begin{APACrefauthors}%
Gelman, A.%
, Carlin, J\BPBI B.%
, Stern, H\BPBI S.%
, Dunson, D\BPBI B.%
, Vehtari, A.%
\BCBL {}\ \BBA {} Rubin, D\BPBI B.%
\end{APACrefauthors}%
\unskip\
\newblock
\APACrefYear{2014}.
\newblock
\APACrefbtitle {{B}ayesian Data Analysis} {{B}ayesian data analysis}\
  (\PrintOrdinal{Third}\ \BEd).
\newblock
\APACaddressPublisher{}{Boca Raton: Taylor \& Francis Group}.
\PrintBackRefs{\CurrentBib}

\bibitem [\protect \citeauthoryear {%
Genz%
\ \protect \BOthers {.}}{%
Genz%
\ \protect \BOthers {.}}{%
{\protect \APACyear {2016}}%
}]{%
Genz:2016}
\APACinsertmetastar {%
Genz:2016}%
\begin{APACrefauthors}%
Genz, A.%
, Bretz, F.%
, Miwa, T.%
, Mi, X.%
, Leisch, F.%
, Scheipl, F.%
\BDBL {}Hothorn, T.%
\end{APACrefauthors}%
\unskip\
\newblock
\APACrefYearMonthDay{2016}{}{}.
\newblock
{\BBOQ}\APACrefatitle {\textbf{mvtnorm}: Multivariate Normal and t
  Distributions} {\textbf{mvtnorm}: Multivariate normal and t
  distributions}{\BBCQ}\ [\bibcomputersoftwaremanual].
\newblock
\begin{APACrefURL}
  \url{https://cran.r-project.org/web/packages/mvtnorm/index.html}
  \end{APACrefURL}
\newblock
\APACrefnote{\textsf{R} package version 1.0-11}
\PrintBackRefs{\CurrentBib}

\bibitem [\protect \citeauthoryear {%
Gronau%
\ \protect \BOthers {.}}{%
Gronau%
\ \protect \BOthers {.}}{%
{\protect \APACyear {2017}}%
}]{%
Gronau:2017}
\APACinsertmetastar {%
Gronau:2017}%
\begin{APACrefauthors}%
Gronau, Q\BPBI F.%
, {van Erp}, S.%
, Heck, D.%
, Cesario, J.%
, Jonas, K.%
\BCBL {}\ \BBA {} Wagenmakers, E\BHBI J.%
\end{APACrefauthors}%
\unskip\
\newblock
\APACrefYearMonthDay{2017}{}{}.
\newblock
{\BBOQ}\APACrefatitle {A Bayesian Model-Averaged Meta-Analysis of the Power
  Pose Effect with Informed and Default Priors: {T}he Case of Felt Power} {A
  bayesian model-averaged meta-analysis of the power pose effect with informed
  and default priors: {T}he case of felt power}.{\BBCQ}
\newblock
\APACjournalVolNumPages{Comprehensive Results in Social
  Psychology}{2}{1}{123--138}.
\PrintBackRefs{\CurrentBib}

\bibitem [\protect \citeauthoryear {%
Gu%
, Hoijtink%
, Mulder%
\BCBL {}\ \BBA {} Rosseel%
}{%
Gu%
\ \protect \BOthers {.}}{%
{\protect \APACyear {2019}}%
}]{%
Gu:2019}
\APACinsertmetastar {%
Gu:2019}%
\begin{APACrefauthors}%
Gu, X.%
, Hoijtink, H.%
, Mulder, J.%
\BCBL {}\ \BBA {} Rosseel, Y.%
\end{APACrefauthors}%
\unskip\
\newblock
\APACrefYearMonthDay{2019}{}{}.
\newblock
{\BBOQ}\APACrefatitle {Bain: A Program for the Evaluation of Inequality
  Constrained Hypotheses Using {B}ayes Factors in Structural Equation Models}
  {Bain: A program for the evaluation of inequality constrained hypotheses
  using {B}ayes factors in structural equation models}.{\BBCQ}
\newblock
\APACjournalVolNumPages{Journal of Statistical Computation and
  Simulation}{}{}{}.
\newblock
\begin{APACrefDOI} \doi{10.1080/00949655.2019.1590574} \end{APACrefDOI}
\PrintBackRefs{\CurrentBib}

\bibitem [\protect \citeauthoryear {%
Gu%
\ \protect \BOthers {.}}{%
Gu%
\ \protect \BOthers {.}}{%
{\protect \APACyear {2018}}%
}]{%
bain}
\APACinsertmetastar {%
bain}%
\begin{APACrefauthors}%
Gu, X.%
, Hoijtink, H.%
, Mulder, J.%
, {van Lissa}, C\BPBI J.%
, Jones, J.%
, Waller, N.%
\BCBL {}\ \BBA {} {The R Core Team}.%
\end{APACrefauthors}%
\unskip\
\newblock
\APACrefYearMonthDay{2018}{}{}.
\newblock
{\BBOQ}\APACrefatitle {\textbf{bain}: Bayes Factors for Informative Hypotheses}
  {\textbf{bain}: Bayes factors for informative hypotheses}{\BBCQ}\
  [\bibcomputersoftwaremanual].
\newblock
\begin{APACrefURL}
  \url{https://cran.r-project.org/web/packages/bain/index.html}
  \end{APACrefURL}
\newblock
\APACrefnote{\textsf{R} package version 0.2.1}
\PrintBackRefs{\CurrentBib}

\bibitem [\protect \citeauthoryear {%
Gu%
, Mulder%
\BCBL {}\ \BBA {} Hoijtink%
}{%
Gu%
\ \protect \BOthers {.}}{%
{\protect \APACyear {2017}}%
}]{%
Gu:2017}
\APACinsertmetastar {%
Gu:2017}%
\begin{APACrefauthors}%
Gu, X.%
, Mulder, J.%
\BCBL {}\ \BBA {} Hoijtink, H.%
\end{APACrefauthors}%
\unskip\
\newblock
\APACrefYearMonthDay{2017}{}{}.
\newblock
{\BBOQ}\APACrefatitle {Approximated Adjusted Fractional Bayes Factors: A
  General Method for Testing Informative Hypotheses} {Approximated adjusted
  fractional bayes factors: A general method for testing informative
  hypotheses}.{\BBCQ}
\newblock
\APACjournalVolNumPages{British Journal of Mathematical and Statistical
  Psychology}{71}{}{229--261}.
\PrintBackRefs{\CurrentBib}

\bibitem [\protect \citeauthoryear {%
Hoijtink%
}{%
Hoijtink%
}{%
{\protect \APACyear {2011}}%
}]{%
Hoijtink:2011}
\APACinsertmetastar {%
Hoijtink:2011}%
\begin{APACrefauthors}%
Hoijtink, H.%
\end{APACrefauthors}%
\unskip\
\newblock
\APACrefYear{2011}.
\newblock
\APACrefbtitle {Informative Hypotheses: Theory and Practice for Behavioral and
  Social Scientists} {Informative hypotheses: Theory and practice for
  behavioral and social scientists}.
\newblock
\APACaddressPublisher{}{New York: Chapman \& Hall/CRC}.
\PrintBackRefs{\CurrentBib}

\bibitem [\protect \citeauthoryear {%
Hoijtink%
\ \BBA {} Chow%
}{%
Hoijtink%
\ \BBA {} Chow%
}{%
{\protect \APACyear {2017}}%
}]{%
Hoijtink:2017}
\APACinsertmetastar {%
Hoijtink:2017}%
\begin{APACrefauthors}%
Hoijtink, H.%
\BCBT {}\ \BBA {} Chow, S\BHBI M.%
\end{APACrefauthors}%
\unskip\
\newblock
\APACrefYearMonthDay{2017}{}{}.
\newblock
{\BBOQ}\APACrefatitle {Bayesian Hypothesis Testing: {E}ditorial to the Special
  Issue on {B}ayesian Data Analysis} {Bayesian hypothesis testing: {E}ditorial
  to the special issue on {B}ayesian data analysis}.{\BBCQ}
\newblock
\APACjournalVolNumPages{Psychological Methods}{22}{2}{211--216}.
\newblock
\begin{APACrefDOI} \doi{10.1037/met0000143} \end{APACrefDOI}
\PrintBackRefs{\CurrentBib}

\bibitem [\protect \citeauthoryear {%
Hoijtink%
, Gu%
\BCBL {}\ \BBA {} Mulder%
}{%
Hoijtink%
, Gu%
\BCBL {}\ \BBA {} Mulder%
}{%
{\protect \APACyear {2018}}%
}]{%
Hoijtink:2018}
\APACinsertmetastar {%
Hoijtink:2018}%
\begin{APACrefauthors}%
Hoijtink, H.%
, Gu, X.%
\BCBL {}\ \BBA {} Mulder, J.%
\end{APACrefauthors}%
\unskip\
\newblock
\APACrefYearMonthDay{2018}{}{}.
\newblock
{\BBOQ}\APACrefatitle {Bayesian Evaluation of Informative Hypotheses for
  Multiple Populations} {Bayesian evaluation of informative hypotheses for
  multiple populations}.{\BBCQ}
\newblock
\APACjournalVolNumPages{British Journal of Mathematical and Statistical
  Psychology}{}{}{}.
\newblock
\begin{APACrefDOI} \doi{doi.org/10.1111/bmsp.12145} \end{APACrefDOI}
\PrintBackRefs{\CurrentBib}

\bibitem [\protect \citeauthoryear {%
Hoijtink%
, Gu%
, Mulder%
\BCBL {}\ \BBA {} Rosseel%
}{%
Hoijtink%
, Gu%
, Mulder%
\BCBL {}\ \BBA {} Rosseel%
}{%
{\protect \APACyear {2018}}%
}]{%
Hoijtink:2018b}
\APACinsertmetastar {%
Hoijtink:2018b}%
\begin{APACrefauthors}%
Hoijtink, H.%
, Gu, X.%
, Mulder, J.%
\BCBL {}\ \BBA {} Rosseel, Y.%
\end{APACrefauthors}%
\unskip\
\newblock
\APACrefYearMonthDay{2018}{}{}.
\newblock
{\BBOQ}\APACrefatitle {Computing {B}ayes Factors From Data With Missing Values}
  {Computing {B}ayes factors from data with missing values}.{\BBCQ}
\newblock
\APACjournalVolNumPages{Psychological Methods}{24}{2}{253--268}.
\newblock
\begin{APACrefDOI} \doi{10.1037/met0000187} \end{APACrefDOI}
\PrintBackRefs{\CurrentBib}

\bibitem [\protect \citeauthoryear {%
Hoijtink%
, Mulder%
, van Lissa%
\BCBL {}\ \BBA {} Gu%
}{%
Hoijtink%
\ \protect \BOthers {.}}{%
{\protect \APACyear {2019}}%
}]{%
Hoijtink:2019}
\APACinsertmetastar {%
Hoijtink:2019}%
\begin{APACrefauthors}%
Hoijtink, H.%
, Mulder, J.%
, van Lissa, C.%
\BCBL {}\ \BBA {} Gu, X.%
\end{APACrefauthors}%
\unskip\
\newblock
\APACrefYearMonthDay{2019}{}{}.
\newblock
{\BBOQ}\APACrefatitle {A Tutorial on Testing Hypotheses Using the {B}ayes
  Factor} {A tutorial on testing hypotheses using the {B}ayes factor}.{\BBCQ}
\newblock
\APACjournalVolNumPages{Psychological Methods}{}{}{}.
\newblock
\begin{APACrefDOI} \doi{10.1037/met0000201} \end{APACrefDOI}
\PrintBackRefs{\CurrentBib}

\bibitem [\protect \citeauthoryear {%
Hothorn%
\ \protect \BOthers {.}}{%
Hothorn%
\ \protect \BOthers {.}}{%
{\protect \APACyear {2019}}%
}]{%
Hothorn:2019}
\APACinsertmetastar {%
Hothorn:2019}%
\begin{APACrefauthors}%
Hothorn, T.%
, Zeileis, A.%
, Farebrother, R\BPBI W.%
, Cummins, C.%
, Millo, G.%
\BCBL {}\ \BBA {} Mitchell, D.%
\end{APACrefauthors}%
\unskip\
\newblock
\APACrefYearMonthDay{2019}{}{}.
\newblock
{\BBOQ}\APACrefatitle {\textbf{lmtest}: Testing Linear Regression Models}
  {\textbf{lmtest}: Testing linear regression models}{\BBCQ}\
  [\bibcomputersoftwaremanual].
\newblock
\begin{APACrefURL}
  \url{https://cran.r-project.org/web/packages/lmtest/index.html}
  \end{APACrefURL}
\newblock
\APACrefnote{\textsf{R} package version 0.9-37}
\PrintBackRefs{\CurrentBib}

\bibitem [\protect \citeauthoryear {%
Howell%
}{%
Howell%
}{%
{\protect \APACyear {2012}}%
}]{%
Howell:2012}
\APACinsertmetastar {%
Howell:2012}%
\begin{APACrefauthors}%
Howell, D.%
\end{APACrefauthors}%
\unskip\
\newblock
\APACrefYear{2012}.
\newblock
\APACrefbtitle {Statistical Methods for Psychology} {Statistical methods for
  psychology}\ (\PrintOrdinal{Eighth}\ \BEd).
\newblock
\APACaddressPublisher{}{Belmont, CA: Cengage Learning}.
\PrintBackRefs{\CurrentBib}

\bibitem [\protect \citeauthoryear {%
Janiszewski%
\ \BBA {} Uy%
}{%
Janiszewski%
\ \BBA {} Uy%
}{%
{\protect \APACyear {2008}}%
}]{%
Janiszewski:2008}
\APACinsertmetastar {%
Janiszewski:2008}%
\begin{APACrefauthors}%
Janiszewski, C.%
\BCBT {}\ \BBA {} Uy, D.%
\end{APACrefauthors}%
\unskip\
\newblock
\APACrefYearMonthDay{2008}{}{}.
\newblock
{\BBOQ}\APACrefatitle {Precision of the Anchor Influences the Amount of
  Adjustment} {Precision of the anchor influences the amount of
  adjustment}.{\BBCQ}
\newblock
\APACjournalVolNumPages{Psychological Science}{19}{2}{121--127}.
\newblock
\begin{APACrefDOI} \doi{10.1111/j.1467-9280.2008.02057.x} \end{APACrefDOI}
\PrintBackRefs{\CurrentBib}

\bibitem [\protect \citeauthoryear {%
Jeffreys%
}{%
Jeffreys%
}{%
{\protect \APACyear {1935}}%
}]{%
Jeffreys:1935}
\APACinsertmetastar {%
Jeffreys:1935}%
\begin{APACrefauthors}%
Jeffreys, H.%
\end{APACrefauthors}%
\unskip\
\newblock
\APACrefYearMonthDay{1935}{}{}.
\newblock
{\BBOQ}\APACrefatitle {Some Tests of Significance, Treated by the Theory of
  Probability} {Some tests of significance, treated by the theory of
  probability}.{\BBCQ}
\newblock
\APACjournalVolNumPages{Proceedings of the Cambridge Philosophy
  Society}{31}{2}{203--222}.
\newblock
\begin{APACrefDOI} \doi{10.1017/S030500410001330X} \end{APACrefDOI}
\PrintBackRefs{\CurrentBib}

\bibitem [\protect \citeauthoryear {%
Jeffreys%
}{%
Jeffreys%
}{%
{\protect \APACyear {1961}}%
}]{%
Jeffreys}
\APACinsertmetastar {%
Jeffreys}%
\begin{APACrefauthors}%
Jeffreys, H.%
\end{APACrefauthors}%
\unskip\
\newblock
\APACrefYear{1961}.
\newblock
\APACrefbtitle {Theory of Probability-3rd ed} {Theory of probability-3rd ed}.
\newblock
\APACaddressPublisher{}{New York: Oxford University Press}.
\PrintBackRefs{\CurrentBib}

\bibitem [\protect \citeauthoryear {%
Joe%
}{%
Joe%
}{%
{\protect \APACyear {2006}}%
}]{%
Joe:2006}
\APACinsertmetastar {%
Joe:2006}%
\begin{APACrefauthors}%
Joe, H.%
\end{APACrefauthors}%
\unskip\
\newblock
\APACrefYearMonthDay{2006}{}{}.
\newblock
{\BBOQ}\APACrefatitle {Generating Random Correlation Matrices Based on Partial
  Correlations} {Generating random correlation matrices based on partial
  correlations}.{\BBCQ}
\newblock
\APACjournalVolNumPages{Journal of Multivariate Analysis}{97}{10}{2177--2189}.
\newblock
\begin{APACrefDOI} \doi{10.1016/j.jmva.2005.05.010} \end{APACrefDOI}
\PrintBackRefs{\CurrentBib}

\bibitem [\protect \citeauthoryear {%
Johnson%
\ \BBA {} Kotz%
}{%
Johnson%
\ \BBA {} Kotz%
}{%
{\protect \APACyear {1970}}%
}]{%
Johnson:1970}
\APACinsertmetastar {%
Johnson:1970}%
\begin{APACrefauthors}%
Johnson, N.%
\BCBT {}\ \BBA {} Kotz, S.%
\end{APACrefauthors}%
\unskip\
\newblock
\APACrefYear{1970}.
\newblock
\APACrefbtitle {Distributions in Statistics: Continuous Univariate
  distributions, Vol. 2} {Distributions in statistics: Continuous univariate
  distributions, vol. 2}.
\newblock
\APACaddressPublisher{}{Boston, MA: Houghton Mifflin Co.}
\PrintBackRefs{\CurrentBib}

\bibitem [\protect \citeauthoryear {%
Klugkist%
, Laudy%
\BCBL {}\ \BBA {} Hoijtink%
}{%
Klugkist%
\ \protect \BOthers {.}}{%
{\protect \APACyear {2005}}%
}]{%
Klugkist:2005}
\APACinsertmetastar {%
Klugkist:2005}%
\begin{APACrefauthors}%
Klugkist, I.%
, Laudy, O.%
\BCBL {}\ \BBA {} Hoijtink, H.%
\end{APACrefauthors}%
\unskip\
\newblock
\APACrefYearMonthDay{2005}{}{}.
\newblock
{\BBOQ}\APACrefatitle {Inequality Constrained Analysis of Variance: A
  {B}ayesian Approach} {Inequality constrained analysis of variance: A
  {B}ayesian approach}.{\BBCQ}
\newblock
\APACjournalVolNumPages{Psychological Methods}{10}{4}{477--493}.
\newblock
\begin{APACrefDOI} \doi{10.1037/1082-989X.10.4.477} \end{APACrefDOI}
\PrintBackRefs{\CurrentBib}

\bibitem [\protect \citeauthoryear {%
Kofler%
\ \protect \BOthers {.}}{%
Kofler%
\ \protect \BOthers {.}}{%
{\protect \APACyear {2013}}%
}]{%
Kofler2013}
\APACinsertmetastar {%
Kofler2013}%
\begin{APACrefauthors}%
Kofler, M\BPBI J.%
, Rapport, M\BPBI D.%
, Sarver, D\BPBI E.%
, Raiker, J\BPBI S.%
, Orban, S\BPBI A.%
, Friedman, L\BPBI M.%
\BCBL {}\ \BBA {} Kolomeyer, E\BPBI G.%
\end{APACrefauthors}%
\unskip\
\newblock
\APACrefYearMonthDay{2013}{}{}.
\newblock
{\BBOQ}\APACrefatitle {Reaction Time Variability in {ADHD}: A Meta-Analytic
  Review of 319 Studies} {Reaction time variability in {ADHD}: A meta-analytic
  review of 319 studies}.{\BBCQ}
\newblock
\APACjournalVolNumPages{Clinical Psychology Review}{33}{6}{795--811}.
\newblock
\begin{APACrefDOI} \doi{10.1016/j.cpr.2013.06.001} \end{APACrefDOI}
\PrintBackRefs{\CurrentBib}

\bibitem [\protect \citeauthoryear {%
Lindley%
}{%
Lindley%
}{%
{\protect \APACyear {1957}}%
}]{%
Lindley:1957}
\APACinsertmetastar {%
Lindley:1957}%
\begin{APACrefauthors}%
Lindley, D\BPBI V.%
\end{APACrefauthors}%
\unskip\
\newblock
\APACrefYearMonthDay{1957}{}{}.
\newblock
{\BBOQ}\APACrefatitle {A Statistical Paradox} {A statistical paradox}.{\BBCQ}
\newblock
\APACjournalVolNumPages{Biometrika}{44}{1}{187--192}.
\PrintBackRefs{\CurrentBib}

\bibitem [\protect \citeauthoryear {%
Little%
\ \BBA {} Rubin%
}{%
Little%
\ \BBA {} Rubin%
}{%
{\protect \APACyear {2002}}%
}]{%
LittleRubin2002}
\APACinsertmetastar {%
LittleRubin2002}%
\begin{APACrefauthors}%
Little, R\BPBI J.%
\BCBT {}\ \BBA {} Rubin, D\BPBI B.%
\end{APACrefauthors}%
\unskip\
\newblock
\APACrefYear{2002}.
\newblock
\APACrefbtitle {Statistical Analysis with Missing Data} {Statistical analysis
  with missing data}.
\newblock
\APACaddressPublisher{New York}{John Wiley}.
\PrintBackRefs{\CurrentBib}

\bibitem [\protect \citeauthoryear {%
Love%
\ \protect \BOthers {.}}{%
Love%
\ \protect \BOthers {.}}{%
{\protect \APACyear {2019}}%
}]{%
JASP2018}
\APACinsertmetastar {%
JASP2018}%
\begin{APACrefauthors}%
Love, J.%
, Selker, R.%
, Marsman, M.%
, Jamil, T.%
, Dropmann, D.%
, Verhagen, J.%
\BDBL {}Wagenmakers, E\BHBI J.%
\end{APACrefauthors}%
\unskip\
\newblock
\APACrefYearMonthDay{2019}{}{}.
\newblock
{\BBOQ}\APACrefatitle {JASP: Graphical Statistical Software for Common
  Statistical Designs} {Jasp: Graphical statistical software for common
  statistical designs}.{\BBCQ}
\newblock
\APACjournalVolNumPages{Journal of Statistical Software}{88}{2}{}.
\newblock
\begin{APACrefDOI} \doi{10.18637/jss.v088.i02} \end{APACrefDOI}
\PrintBackRefs{\CurrentBib}

\bibitem [\protect \citeauthoryear {%
Marin%
\ \BBA {} Robert%
}{%
Marin%
\ \BBA {} Robert%
}{%
{\protect \APACyear {2010}}%
}]{%
Marin:2010}
\APACinsertmetastar {%
Marin:2010}%
\begin{APACrefauthors}%
Marin, J\BHBI M.%
\BCBT {}\ \BBA {} Robert, C.%
\end{APACrefauthors}%
\unskip\
\newblock
\APACrefYearMonthDay{2010}{}{}.
\newblock
{\BBOQ}\APACrefatitle {On Resolving the {S}avage-{D}ickey Paradox} {On
  resolving the {S}avage-{D}ickey paradox}.{\BBCQ}
\newblock
\APACjournalVolNumPages{Electronic Journal of Statistics}{4}{}{643--654}.
\newblock
\begin{APACrefDOI} \doi{10.1214/10-EJS564} \end{APACrefDOI}
\PrintBackRefs{\CurrentBib}

\bibitem [\protect \citeauthoryear {%
Masson%
}{%
Masson%
}{%
{\protect \APACyear {2011}}%
}]{%
Masson:2011}
\APACinsertmetastar {%
Masson:2011}%
\begin{APACrefauthors}%
Masson, M.%
\end{APACrefauthors}%
\unskip\
\newblock
\APACrefYearMonthDay{2011}{}{}.
\newblock
{\BBOQ}\APACrefatitle {A Tutorial on a Practical {B}ayesian Alternative to
  Null-Hypothesis Significance Testing} {A tutorial on a practical {B}ayesian
  alternative to null-hypothesis significance testing}.{\BBCQ}
\newblock
\APACjournalVolNumPages{Behavior Research Methods}{43}{3}{679--690}.
\newblock
\begin{APACrefDOI} \doi{10.3758/s13428-010-0049-5} \end{APACrefDOI}
\PrintBackRefs{\CurrentBib}

\bibitem [\protect \citeauthoryear {%
McGuigin%
\ \protect \BOthers {.}}{%
McGuigin%
\ \protect \BOthers {.}}{%
{\protect \APACyear {under review}}%
}]{%
McGuigin:2019}
\APACinsertmetastar {%
McGuigin:2019}%
\begin{APACrefauthors}%
McGuigin, R.%
, A.T.%
, Newton%
, {Tamber-Rosenau}, B.%
, Tomarken, A.%
\BCBL {}\ \BBA {} Gauthier, I.%
\end{APACrefauthors}%
\unskip\
\newblock
\APACrefYearMonthDay{under review}{}{}.
\newblock
{\BBOQ}\APACrefatitle {Thickness of Deep Layers in the Fusiform Face Area
  Predicts Face Recognition} {Thickness of deep layers in the fusiform face
  area predicts face recognition}.{\BBCQ}
\newblock

\PrintBackRefs{\CurrentBib}

\bibitem [\protect \citeauthoryear {%
McGuigin%
, {Van Gulick}%
\BCBL {}\ \BBA {} Gauthier%
}{%
McGuigin%
\ \protect \BOthers {.}}{%
{\protect \APACyear {2016}}%
}]{%
McGuigin:2016}
\APACinsertmetastar {%
McGuigin:2016}%
\begin{APACrefauthors}%
McGuigin, R.%
, {Van Gulick}, A.%
\BCBL {}\ \BBA {} Gauthier, I.%
\end{APACrefauthors}%
\unskip\
\newblock
\APACrefYearMonthDay{2016}{}{}.
\newblock
{\BBOQ}\APACrefatitle {Cortical Thickness in Fusiform Face Area Predicts Face
  and Object Recognition Performance} {Cortical thickness in fusiform face area
  predicts face and object recognition performance}.{\BBCQ}
\newblock
\APACjournalVolNumPages{Journal of Cognitive Neuroscience}{28}{2}{282--294}.
\newblock
\begin{APACrefDOI} \doi{10.1162/jocn_a_00891} \end{APACrefDOI}
\PrintBackRefs{\CurrentBib}

\bibitem [\protect \citeauthoryear {%
Morey%
\ \protect \BOthers {.}}{%
Morey%
\ \protect \BOthers {.}}{%
{\protect \APACyear {2018}}%
}]{%
BayesFactor}
\APACinsertmetastar {%
BayesFactor}%
\begin{APACrefauthors}%
Morey, R.%
, Rouder, J\BPBI N.%
, Jamil, T.%
, Urbanek, S.%
, Forner, K.%
\BCBL {}\ \BBA {} Ly, A.%
\end{APACrefauthors}%
\unskip\
\newblock
\APACrefYearMonthDay{2018}{}{}.
\newblock
\APACrefbtitle {\textbf{BayesFactor}: Computation of Bayes Factors for Common
  Designs.} {\textbf{BayesFactor}: Computation of bayes factors for common
  designs.}
\newblock
\begin{APACrefURL}
  \url{https://cran.r-project.org/web/packages/BayesFactor/index.html}
  \end{APACrefURL}
\newblock
\APACrefnote{\textsf{R} package version 0.9.12-4.2}
\PrintBackRefs{\CurrentBib}

\bibitem [\protect \citeauthoryear {%
Mulder%
}{%
Mulder%
}{%
{\protect \APACyear {2014}}%
{\protect \APACexlab {{\protect \BCnt {1}}}}}]{%
Mulder:2014a}
\APACinsertmetastar {%
Mulder:2014a}%
\begin{APACrefauthors}%
Mulder, J.%
\end{APACrefauthors}%
\unskip\
\newblock
\APACrefYearMonthDay{2014{\protect \BCnt {1}}}{}{}.
\newblock
{\BBOQ}\APACrefatitle {{B}ayes Factors for Testing Inequality Constrained
  Hypotheses: Issues with Prior Specification} {{B}ayes factors for testing
  inequality constrained hypotheses: Issues with prior specification}.{\BBCQ}
\newblock
\APACjournalVolNumPages{British Journal of Statistical and Mathematical
  Psychology}{67}{1}{153--71}.
\newblock
\begin{APACrefDOI} \doi{10.1111/bmsp.12013} \end{APACrefDOI}
\PrintBackRefs{\CurrentBib}

\bibitem [\protect \citeauthoryear {%
Mulder%
}{%
Mulder%
}{%
{\protect \APACyear {2014}}%
{\protect \APACexlab {{\protect \BCnt {2}}}}}]{%
Mulder:2014b}
\APACinsertmetastar {%
Mulder:2014b}%
\begin{APACrefauthors}%
Mulder, J.%
\end{APACrefauthors}%
\unskip\
\newblock
\APACrefYearMonthDay{2014{\protect \BCnt {2}}}{}{}.
\newblock
{\BBOQ}\APACrefatitle {Prior Adjusted Default {B}ayes Factors for Testing
  (In)equality Constrained Hypotheses} {Prior adjusted default {B}ayes factors
  for testing (in)equality constrained hypotheses}.{\BBCQ}
\newblock
\APACjournalVolNumPages{Computational Statistics and Data
  Analysis}{71}{}{448--463}.
\newblock
\begin{APACrefDOI} \doi{10.1016/j.csda.2013.07.017} \end{APACrefDOI}
\PrintBackRefs{\CurrentBib}

\bibitem [\protect \citeauthoryear {%
Mulder%
}{%
Mulder%
}{%
{\protect \APACyear {2016}}%
}]{%
Mulder:2016}
\APACinsertmetastar {%
Mulder:2016}%
\begin{APACrefauthors}%
Mulder, J.%
\end{APACrefauthors}%
\unskip\
\newblock
\APACrefYearMonthDay{2016}{}{}.
\newblock
{\BBOQ}\APACrefatitle {Bayes Factors for Testing Order-Constrained Hypotheses
  on Correlations} {Bayes factors for testing order-constrained hypotheses on
  correlations}.{\BBCQ}
\newblock
\APACjournalVolNumPages{Journal of Mathematical Psychology}{72}{}{104-115}.
\newblock
\begin{APACrefDOI} \doi{10.1016/j.jmp.2014.09.004} \end{APACrefDOI}
\PrintBackRefs{\CurrentBib}

\bibitem [\protect \citeauthoryear {%
Mulder%
\ \BBA {} Fox%
}{%
Mulder%
\ \BBA {} Fox%
}{%
{\protect \APACyear {2013}}%
}]{%
MulderFox:2013}
\APACinsertmetastar {%
MulderFox:2013}%
\begin{APACrefauthors}%
Mulder, J.%
\BCBT {}\ \BBA {} Fox, J\BHBI P.%
\end{APACrefauthors}%
\unskip\
\newblock
\APACrefYearMonthDay{2013}{}{}.
\newblock
{\BBOQ}\APACrefatitle {Bayesian Tests on Components of the Compound Symmetry
  Covariance Matrix} {Bayesian tests on components of the compound symmetry
  covariance matrix}.{\BBCQ}
\newblock
\APACjournalVolNumPages{Statistics and Computing}{23}{}{109--122}.
\newblock
\begin{APACrefDOI} \doi{10.1007/s11222-011-9295-3} \end{APACrefDOI}
\PrintBackRefs{\CurrentBib}

\bibitem [\protect \citeauthoryear {%
Mulder%
\ \BBA {} Fox%
}{%
Mulder%
\ \BBA {} Fox%
}{%
{\protect \APACyear {2019}}%
}]{%
MulderFox:2019}
\APACinsertmetastar {%
MulderFox:2019}%
\begin{APACrefauthors}%
Mulder, J.%
\BCBT {}\ \BBA {} Fox, J\BHBI P.%
\end{APACrefauthors}%
\unskip\
\newblock
\APACrefYearMonthDay{2019}{}{}.
\newblock
{\BBOQ}\APACrefatitle {Bayes Factor Testing of Multiple Intraclass
  Correlations} {Bayes factor testing of multiple intraclass
  correlations}.{\BBCQ}
\newblock
\APACjournalVolNumPages{Bayesian Analysis}{14}{}{521--552}.
\PrintBackRefs{\CurrentBib}

\bibitem [\protect \citeauthoryear {%
Mulder%
\ \BBA {} Gelissen%
}{%
Mulder%
\ \BBA {} Gelissen%
}{%
{\protect \APACyear {2019}}%
}]{%
MulderGelissen:2019}
\APACinsertmetastar {%
MulderGelissen:2019}%
\begin{APACrefauthors}%
Mulder, J.%
\BCBT {}\ \BBA {} Gelissen, J.%
\end{APACrefauthors}%
\unskip\
\newblock
\APACrefYearMonthDay{2019}{}{}.
\newblock
{\BBOQ}\APACrefatitle {Bayes Factor Testing of Equality and Order Constraints
  on Measures of Association in Social Research} {Bayes factor testing of
  equality and order constraints on measures of association in social
  research}.{\BBCQ}
\newblock
\begin{APACrefURL} \url{https://arxiv.org/abs/1807.05819} \end{APACrefURL}
\PrintBackRefs{\CurrentBib}

\bibitem [\protect \citeauthoryear {%
Mulder%
, Hoijtink%
\BCBL {}\ \BBA {} de Leeuw%
}{%
Mulder%
\ \protect \BOthers {.}}{%
{\protect \APACyear {2012}}%
}]{%
Mulder:2012}
\APACinsertmetastar {%
Mulder:2012}%
\begin{APACrefauthors}%
Mulder, J.%
, Hoijtink, H.%
\BCBL {}\ \BBA {} de Leeuw, C.%
\end{APACrefauthors}%
\unskip\
\newblock
\APACrefYearMonthDay{2012}{}{}.
\newblock
{\BBOQ}\APACrefatitle {BIEMS: A Fortran 90 Program for Calculating {B}ayes
  Factors for Inequality and Equality Constrained Model} {Biems: A fortran 90
  program for calculating {B}ayes factors for inequality and equality
  constrained model}.{\BBCQ}
\newblock
\APACjournalVolNumPages{Journal of Statistical Software}{46}{}{}.
\newblock
\begin{APACrefDOI} \doi{10.18637/jss.v046.i02} \end{APACrefDOI}
\PrintBackRefs{\CurrentBib}

\bibitem [\protect \citeauthoryear {%
Mulder%
, Hoijtink%
\BCBL {}\ \BBA {} Gu%
}{%
Mulder%
\ \protect \BOthers {.}}{%
{\protect \APACyear {2019}}%
}]{%
Mulder:2019}
\APACinsertmetastar {%
Mulder:2019}%
\begin{APACrefauthors}%
Mulder, J.%
, Hoijtink, H.%
\BCBL {}\ \BBA {} Gu, X.%
\end{APACrefauthors}%
\unskip\
\newblock
\APACrefYearMonthDay{2019}{}{}.
\newblock
{\BBOQ}\APACrefatitle {Default {B}ayesian Model Selection of Constrained
  Multivariate Normal Linear Models} {Default {B}ayesian model selection of
  constrained multivariate normal linear models}.{\BBCQ}
\newblock
\begin{APACrefURL} \url{https://arxiv.org/abs/1904.00679} \end{APACrefURL}
\PrintBackRefs{\CurrentBib}

\bibitem [\protect \citeauthoryear {%
Mulder%
, Hoijtink%
\BCBL {}\ \BBA {} Klugkist%
}{%
Mulder%
\ \protect \BOthers {.}}{%
{\protect \APACyear {2010}}%
}]{%
Mulder:2010}
\APACinsertmetastar {%
Mulder:2010}%
\begin{APACrefauthors}%
Mulder, J.%
, Hoijtink, H.%
\BCBL {}\ \BBA {} Klugkist, I.%
\end{APACrefauthors}%
\unskip\
\newblock
\APACrefYearMonthDay{2010}{}{}.
\newblock
{\BBOQ}\APACrefatitle {Equality and Inequality Constrained Multivariate Linear
  Models: Objective Model Selection Using Constrained Posterior Priors}
  {Equality and inequality constrained multivariate linear models: Objective
  model selection using constrained posterior priors}.{\BBCQ}
\newblock
\APACjournalVolNumPages{Journal of Statistical Planning and
  Inference}{140}{4}{887--906}.
\newblock
\begin{APACrefDOI} \doi{10.1016/j.jspi.2009.09.022} \end{APACrefDOI}
\PrintBackRefs{\CurrentBib}

\bibitem [\protect \citeauthoryear {%
Mulder%
\ \BBA {} Olsson-Collentine%
}{%
Mulder%
\ \BBA {} Olsson-Collentine%
}{%
{\protect \APACyear {2019}}%
}]{%
MulderOlsson:2019}
\APACinsertmetastar {%
MulderOlsson:2019}%
\begin{APACrefauthors}%
Mulder, J.%
\BCBT {}\ \BBA {} Olsson-Collentine, A.%
\end{APACrefauthors}%
\unskip\
\newblock
\APACrefYearMonthDay{2019}{}{}.
\newblock
{\BBOQ}\APACrefatitle {Simple Bayesian Testing of Scientific Expectations in
  Linear Regression Models} {Simple bayesian testing of scientific expectations
  in linear regression models}.{\BBCQ}
\newblock
\APACjournalVolNumPages{Behavioral Research Methods}{51}{3}{1117--1130}.
\newblock
\begin{APACrefDOI} \doi{10.3758/s13428-018-01196-9} \end{APACrefDOI}
\PrintBackRefs{\CurrentBib}

\bibitem [\protect \citeauthoryear {%
Mulder%
\ \BBA {} Raftery%
}{%
Mulder%
\ \BBA {} Raftery%
}{%
{\protect \APACyear {accepted}}%
}]{%
MulderRaftery:2019}
\APACinsertmetastar {%
MulderRaftery:2019}%
\begin{APACrefauthors}%
Mulder, J.%
\BCBT {}\ \BBA {} Raftery, A.%
\end{APACrefauthors}%
\unskip\
\newblock
\APACrefYearMonthDay{accepted}{}{}.
\newblock
{\BBOQ}\APACrefatitle {BIC Extensions for Order-Constrained Model Selection}
  {Bic extensions for order-constrained model selection}.{\BBCQ}
\newblock
\APACjournalVolNumPages{Sociological Methods \& Research}{}{}{}.
\newblock
\begin{APACrefURL} \url{https://arxiv.org/abs/1805.10639} \end{APACrefURL}
\PrintBackRefs{\CurrentBib}

\bibitem [\protect \citeauthoryear {%
Mulder%
\ \BBA {} Wagenmakers%
}{%
Mulder%
\ \BBA {} Wagenmakers%
}{%
{\protect \APACyear {2016}}%
}]{%
MulderWagenmakers:2016}
\APACinsertmetastar {%
MulderWagenmakers:2016}%
\begin{APACrefauthors}%
Mulder, J.%
\BCBT {}\ \BBA {} Wagenmakers, E\BHBI J.%
\end{APACrefauthors}%
\unskip\
\newblock
\APACrefYearMonthDay{2016}{}{}.
\newblock
{\BBOQ}\APACrefatitle {Editors' Introduction to the Special Issue ``Bayes
  Factors for Testing Hypotheses in Psychological Research: Practical Relevance
  and New Developments''} {Editors' introduction to the special issue ``bayes
  factors for testing hypotheses in psychological research: Practical relevance
  and new developments''}.{\BBCQ}
\newblock
\APACjournalVolNumPages{Journal of Mathematical Psychology}{72}{}{1--5}.
\newblock
\begin{APACrefDOI} \doi{10.1016/j.jmp.2016.01.002} \end{APACrefDOI}
\PrintBackRefs{\CurrentBib}

\bibitem [\protect \citeauthoryear {%
O'Hagan%
}{%
O'Hagan%
}{%
{\protect \APACyear {1995}}%
}]{%
OHagan:1995}
\APACinsertmetastar {%
OHagan:1995}%
\begin{APACrefauthors}%
O'Hagan, A.%
\end{APACrefauthors}%
\unskip\
\newblock
\APACrefYearMonthDay{1995}{}{}.
\newblock
{\BBOQ}\APACrefatitle {Fractional {B}ayes Factors for Model Comparison (with
  discussion)} {Fractional {B}ayes factors for model comparison (with
  discussion)}.{\BBCQ}
\newblock
\APACjournalVolNumPages{Journal of the Royal Statistical Society
  B}{57}{1}{99--138}.
\newblock
\begin{APACrefDOI} \doi{10.1111/j.2517-6161.1995.tb02017.x} \end{APACrefDOI}
\PrintBackRefs{\CurrentBib}

\bibitem [\protect \citeauthoryear {%
O'Hagan%
}{%
O'Hagan%
}{%
{\protect \APACyear {1997}}%
}]{%
OHagan:1997}
\APACinsertmetastar {%
OHagan:1997}%
\begin{APACrefauthors}%
O'Hagan, A.%
\end{APACrefauthors}%
\unskip\
\newblock
\APACrefYearMonthDay{1997}{}{}.
\newblock
{\BBOQ}\APACrefatitle {Properties of Intrinsic and Fractional {B}ayes Factors}
  {Properties of intrinsic and fractional {B}ayes factors}.{\BBCQ}
\newblock
\APACjournalVolNumPages{Test}{6}{1}{101--118}.
\newblock
\begin{APACrefDOI} \doi{10.1007/BF02564428} \end{APACrefDOI}
\PrintBackRefs{\CurrentBib}

\bibitem [\protect \citeauthoryear {%
Pericchi%
, Liu%
\BCBL {}\ \BBA {} Torres%
}{%
Pericchi%
\ \protect \BOthers {.}}{%
{\protect \APACyear {2008}}%
}]{%
Pericchi:2008}
\APACinsertmetastar {%
Pericchi:2008}%
\begin{APACrefauthors}%
Pericchi, L\BPBI R.%
, Liu, G.%
\BCBL {}\ \BBA {} Torres, D.%
\end{APACrefauthors}%
\unskip\
\newblock
\APACrefYearMonthDay{2008}{}{}.
\newblock
{\BBOQ}\APACrefatitle {Objective Bayes Factors for Informative Hypotheses:
  ``Completing'' the Informative Hypothesis and ``Splitting'' the Bayes
  Factors} {Objective bayes factors for informative hypotheses: ``completing''
  the informative hypothesis and ``splitting'' the bayes factors}.{\BBCQ}
\newblock
\BIn{} H.~Hoijtink, I.~Klugkist\BCBL {}\ \BBA {} P\BPBI A.~Boelen\ (\BEDS),
  \APACrefbtitle {Bayesian Evaluation of Informative Hypotheses} {Bayesian
  evaluation of informative hypotheses}\ (\BPGS\ 131--154).
\newblock
\APACaddressPublisher{}{New York: Springer-Verlag}.
\PrintBackRefs{\CurrentBib}

\bibitem [\protect \citeauthoryear {%
{R Development Core Team}%
}{%
{R Development Core Team}%
}{%
{\protect \APACyear {2013}}%
}]{%
Rdevelopment}
\APACinsertmetastar {%
Rdevelopment}%
\begin{APACrefauthors}%
{R Development Core Team}.%
\end{APACrefauthors}%
\unskip\
\newblock
\APACrefYearMonthDay{2013}{}{}.
\newblock
{\BBOQ}\APACrefatitle {\textsf{R}: A Language and Environment for Statistical
  Computing} {\textsf{R}: A language and environment for statistical
  computing}{\BBCQ}\ [\bibcomputersoftwaremanual].
\newblock
\APACaddressPublisher{Vienna, Austria}{}.
\newblock
\begin{APACrefURL} \url{http://www.R-project.org/} \end{APACrefURL}
\PrintBackRefs{\CurrentBib}

\bibitem [\protect \citeauthoryear {%
Rosa%
, Rosa%
, Sarner%
\BCBL {}\ \BBA {} Barrett%
}{%
Rosa%
\ \protect \BOthers {.}}{%
{\protect \APACyear {1998}}%
}]{%
Rosa:1998}
\APACinsertmetastar {%
Rosa:1998}%
\begin{APACrefauthors}%
Rosa, L.%
, Rosa, E.%
, Sarner, L.%
\BCBL {}\ \BBA {} Barrett, S.%
\end{APACrefauthors}%
\unskip\
\newblock
\APACrefYearMonthDay{1998}{}{}.
\newblock
{\BBOQ}\APACrefatitle {A Close Look at Therapeutic Touch} {A close look at
  therapeutic touch}.{\BBCQ}
\newblock
\APACjournalVolNumPages{Journal of the American Medical
  Association}{279}{13}{1005--1010}.
\newblock
\begin{APACrefDOI} \doi{10.1001/jama.279.13.1005} \end{APACrefDOI}
\PrintBackRefs{\CurrentBib}

\bibitem [\protect \citeauthoryear {%
Rouder%
, Speckman%
, D.~Sun%
\BCBL {}\ \BBA {} Iverson%
}{%
Rouder%
\ \protect \BOthers {.}}{%
{\protect \APACyear {2009}}%
}]{%
Rouder:2009}
\APACinsertmetastar {%
Rouder:2009}%
\begin{APACrefauthors}%
Rouder, J\BPBI N.%
, Speckman, P\BPBI L.%
, D.~Sun, R\BPBI D\BPBI M.%
\BCBL {}\ \BBA {} Iverson, G.%
\end{APACrefauthors}%
\unskip\
\newblock
\APACrefYearMonthDay{2009}{}{}.
\newblock
{\BBOQ}\APACrefatitle {Bayesian t Tests for Accepting and Rejecting the Null
  Hypothesis} {Bayesian t tests for accepting and rejecting the null
  hypothesis}.{\BBCQ}
\newblock
\APACjournalVolNumPages{Psychonomic Bulletin \& Review}{16}{2}{225--237}.
\newblock
\begin{APACrefDOI} \doi{10.3758/PBR.16.2.225} \end{APACrefDOI}
\PrintBackRefs{\CurrentBib}

\bibitem [\protect \citeauthoryear {%
Rousseeuw%
\ \BBA {} Molenberghs%
}{%
Rousseeuw%
\ \BBA {} Molenberghs%
}{%
{\protect \APACyear {1994}}%
}]{%
Rousseeuw:1994}
\APACinsertmetastar {%
Rousseeuw:1994}%
\begin{APACrefauthors}%
Rousseeuw, P\BPBI J.%
\BCBT {}\ \BBA {} Molenberghs, G.%
\end{APACrefauthors}%
\unskip\
\newblock
\APACrefYearMonthDay{1994}{}{}.
\newblock
{\BBOQ}\APACrefatitle {The Shape of Correlation Matrices} {The shape of
  correlation matrices}.{\BBCQ}
\newblock
\APACjournalVolNumPages{The American Statistician}{48}{4}{276--279}.
\newblock
\begin{APACrefDOI} \doi{10.2307/2684832} \end{APACrefDOI}
\PrintBackRefs{\CurrentBib}

\bibitem [\protect \citeauthoryear {%
Rubin%
}{%
Rubin%
}{%
{\protect \APACyear {1987}}%
}]{%
Rubin:1987}
\APACinsertmetastar {%
Rubin:1987}%
\begin{APACrefauthors}%
Rubin, D\BPBI B.%
\end{APACrefauthors}%
\unskip\
\newblock
\APACrefYear{1987}.
\newblock
\APACrefbtitle {Multiple Imputation for Nonresponse in Surveys} {Multiple
  imputation for nonresponse in surveys}.
\newblock
\APACaddressPublisher{}{New York: John Wiley}.
\PrintBackRefs{\CurrentBib}

\bibitem [\protect \citeauthoryear {%
Rubin%
}{%
Rubin%
}{%
{\protect \APACyear {1996}}%
}]{%
Rubin:1996}
\APACinsertmetastar {%
Rubin:1996}%
\begin{APACrefauthors}%
Rubin, D\BPBI B.%
\end{APACrefauthors}%
\unskip\
\newblock
\APACrefYearMonthDay{1996}{}{}.
\newblock
{\BBOQ}\APACrefatitle {Multiple Imputation After 18+ Years} {Multiple
  imputation after 18+ years}.{\BBCQ}
\newblock
\APACjournalVolNumPages{Journal of the American statistical
  Association}{91}{434}{473--489}.
\newblock
\begin{APACrefDOI} \doi{10.2307/2291635} \end{APACrefDOI}
\PrintBackRefs{\CurrentBib}

\bibitem [\protect \citeauthoryear {%
Russell%
\ \protect \BOthers {.}}{%
Russell%
\ \protect \BOthers {.}}{%
{\protect \APACyear {2006}}%
}]{%
Russell2006}
\APACinsertmetastar {%
Russell2006}%
\begin{APACrefauthors}%
Russell, V\BPBI A.%
, Oades, R\BPBI D.%
, Tannock, R.%
, Killeen, P\BPBI R.%
, Auerbach, J\BPBI G.%
, Johansen, E\BPBI B.%
\BCBL {}\ \BBA {} Sagvolden, T.%
\end{APACrefauthors}%
\unskip\
\newblock
\APACrefYearMonthDay{2006}{}{}.
\newblock
{\BBOQ}\APACrefatitle {Response Variability in {Attention-Deficit/Hyperactivity
  Disorder}: A Neuronal and Glial Energetics Hypothesis} {Response variability
  in {Attention-Deficit/Hyperactivity Disorder}: A neuronal and glial
  energetics hypothesis}.{\BBCQ}
\newblock
\APACjournalVolNumPages{Behavioral and Brain Functions}{2}{1}{1--25}.
\newblock
\begin{APACrefDOI} \doi{10.1186/1744-9081-2-30} \end{APACrefDOI}
\PrintBackRefs{\CurrentBib}

\bibitem [\protect \citeauthoryear {%
Saville%
\ \BBA {} Herring%
}{%
Saville%
\ \BBA {} Herring%
}{%
{\protect \APACyear {2009}}%
}]{%
Saville:2009}
\APACinsertmetastar {%
Saville:2009}%
\begin{APACrefauthors}%
Saville, B.%
\BCBT {}\ \BBA {} Herring, A.%
\end{APACrefauthors}%
\unskip\
\newblock
\APACrefYearMonthDay{2009}{}{}.
\newblock
{\BBOQ}\APACrefatitle {Testing Random Effects in the Linear Mixed Model Using
  Approximate Bayes Factors} {Testing random effects in the linear mixed model
  using approximate bayes factors}.{\BBCQ}
\newblock
\APACjournalVolNumPages{Biometrics}{65}{2}{369--376}.
\newblock
\begin{APACrefDOI} \doi{10.1111/j.1541-0420.2008.01107.x} \end{APACrefDOI}
\PrintBackRefs{\CurrentBib}

\bibitem [\protect \citeauthoryear {%
Sch\"{o}nbrodt%
, Wagenmakers%
, Zehetleitner%
\BCBL {}\ \BBA {} Perugini%
}{%
Sch\"{o}nbrodt%
\ \protect \BOthers {.}}{%
{\protect \APACyear {2017}}%
}]{%
Schonbrodt:2017}
\APACinsertmetastar {%
Schonbrodt:2017}%
\begin{APACrefauthors}%
Sch\"{o}nbrodt, F\BPBI D.%
, Wagenmakers, E\BHBI J.%
, Zehetleitner, M.%
\BCBL {}\ \BBA {} Perugini, M.%
\end{APACrefauthors}%
\unskip\
\newblock
\APACrefYearMonthDay{2017}{}{}.
\newblock
{\BBOQ}\APACrefatitle {Sequential Hypothesis Testing with {B}ayes Factors:
  {E}fficiently Testing Mean Differences} {Sequential hypothesis testing with
  {B}ayes factors: {E}fficiently testing mean differences}.{\BBCQ}
\newblock
\APACjournalVolNumPages{Psychological Methods}{22}{2}{322--339}.
\newblock
\begin{APACrefDOI} \doi{10.1037/met0000061} \end{APACrefDOI}
\PrintBackRefs{\CurrentBib}

\bibitem [\protect \citeauthoryear {%
Sellke%
, Bayarri%
\BCBL {}\ \BBA {} Berger%
}{%
Sellke%
\ \protect \BOthers {.}}{%
{\protect \APACyear {2001}}%
}]{%
Sellke:2001}
\APACinsertmetastar {%
Sellke:2001}%
\begin{APACrefauthors}%
Sellke, T.%
, Bayarri, M\BPBI J.%
\BCBL {}\ \BBA {} Berger, J\BPBI O.%
\end{APACrefauthors}%
\unskip\
\newblock
\APACrefYearMonthDay{2001}{}{}.
\newblock
{\BBOQ}\APACrefatitle {Calibration of $p$ Values for Testing Precise Null
  Hypotheses} {Calibration of $p$ values for testing precise null
  hypotheses}.{\BBCQ}
\newblock
\APACjournalVolNumPages{The American Statistician}{55}{1}{62--71}.
\newblock
\begin{APACrefDOI} \doi{10.1198/000313001300339950} \end{APACrefDOI}
\PrintBackRefs{\CurrentBib}

\bibitem [\protect \citeauthoryear {%
Silverstein%
, Como%
, Palumbo%
, West%
\BCBL {}\ \BBA {} Osborn%
}{%
Silverstein%
\ \protect \BOthers {.}}{%
{\protect \APACyear {1995}}%
}]{%
Silverstein1995}
\APACinsertmetastar {%
Silverstein1995}%
\begin{APACrefauthors}%
Silverstein, S\BPBI M.%
, Como, P\BPBI G.%
, Palumbo, D\BPBI R.%
, West, L\BPBI L.%
\BCBL {}\ \BBA {} Osborn, L\BPBI M.%
\end{APACrefauthors}%
\unskip\
\newblock
\APACrefYearMonthDay{1995}{}{}.
\newblock
{\BBOQ}\APACrefatitle {Multiple Sources of Attentional Dysfunction in Adults
  With {Tourette's} Syndrome: Comparison With Attention Deficit-Hyperactivity
  Disorder} {Multiple sources of attentional dysfunction in adults with
  {Tourette's} syndrome: Comparison with attention deficit-hyperactivity
  disorder}.{\BBCQ}
\newblock
\APACjournalVolNumPages{Neuropsychology}{9}{2}{157--164}.
\PrintBackRefs{\CurrentBib}

\bibitem [\protect \citeauthoryear {%
Thalmann%
, Niklaus%
\BCBL {}\ \BBA {} Oberauer%
}{%
Thalmann%
\ \protect \BOthers {.}}{%
{\protect \APACyear {2017}}%
}]{%
Thalmann:2017}
\APACinsertmetastar {%
Thalmann:2017}%
\begin{APACrefauthors}%
Thalmann, M.%
, Niklaus, M.%
\BCBL {}\ \BBA {} Oberauer, K.%
\end{APACrefauthors}%
\unskip\
\newblock
\APACrefYearMonthDay{2017}{}{}.
\newblock
{\BBOQ}\APACrefatitle {Estimating Bayes Factors for Linear Models and Random
  Slopes and Continuous Predictors} {Estimating bayes factors for linear models
  and random slopes and continuous predictors}.{\BBCQ}
\newblock

\PrintBackRefs{\CurrentBib}

\bibitem [\protect \citeauthoryear {%
{van Buuren}%
\ \protect \BOthers {.}}{%
{van Buuren}%
\ \protect \BOthers {.}}{%
{\protect \APACyear {2019}}%
}]{%
mice:2019}
\APACinsertmetastar {%
mice:2019}%
\begin{APACrefauthors}%
{van Buuren}, S.%
, {Groothuis-Oudshoorn}, K.%
, Robitzsch, A.%
, Vink, G.%
, Doove, L.%
, Jolani, S.%
\BDBL {}Gray, B.%
\end{APACrefauthors}%
\unskip\
\newblock
\APACrefYearMonthDay{2019}{}{}.
\newblock
{\BBOQ}\APACrefatitle {\textbf{mice}: Multivariate Imputation by Chained
  Equations} {\textbf{mice}: Multivariate imputation by chained
  equations}{\BBCQ}\ [\bibcomputersoftwaremanual].
\newblock
\begin{APACrefURL}
  \url{https://cran.r-project.org/web/packages/mice/index.html}
  \end{APACrefURL}
\newblock
\APACrefnote{\textsf{R} package version 3.5.0}
\PrintBackRefs{\CurrentBib}

\bibitem [\protect \citeauthoryear {%
{Van de Schoot}%
\ \protect \BOthers {.}}{%
{Van de Schoot}%
\ \protect \BOthers {.}}{%
{\protect \APACyear {2006}}%
}]{%
Schoot:2011c}
\APACinsertmetastar {%
Schoot:2011c}%
\begin{APACrefauthors}%
{Van de Schoot}, R.%
, Hoijtink, H.%
, Mulder, J.%
, Aken, M.%
, {de Castro}, B.%
, Meeus, W.%
\BCBL {}\ \BBA {} Romeijn, J\BHBI W.%
\end{APACrefauthors}%
\unskip\
\newblock
\APACrefYearMonthDay{2006}{}{}.
\newblock
{\BBOQ}\APACrefatitle {Evaluating Expectations About Negative Emotional States
  of Aggressive Boys Using {B}ayesian Model Selection} {Evaluating expectations
  about negative emotional states of aggressive boys using {B}ayesian model
  selection}.{\BBCQ}
\newblock
\APACjournalVolNumPages{Developmental Psychology}{47}{}{203--212}.
\newblock
\begin{APACrefDOI} \doi{10.1037/a0020957} \end{APACrefDOI}
\PrintBackRefs{\CurrentBib}

\bibitem [\protect \citeauthoryear {%
{van Ravenzwaaij}%
, Monden%
, Tendeiro%
\BCBL {}\ \BBA {} Ioannidis%
}{%
{van Ravenzwaaij}%
\ \protect \BOthers {.}}{%
{\protect \APACyear {2019}}%
}]{%
Raavenswaaij:2019}
\APACinsertmetastar {%
Raavenswaaij:2019}%
\begin{APACrefauthors}%
{van Ravenzwaaij}, D.%
, Monden, R.%
, Tendeiro, J.%
\BCBL {}\ \BBA {} Ioannidis, J.%
\end{APACrefauthors}%
\unskip\
\newblock
\APACrefYearMonthDay{2019}{}{}.
\newblock
{\BBOQ}\APACrefatitle {Bayes Factors for Superiority, Non-Inferiority, and
  Equivalence Designs} {Bayes factors for superiority, non-inferiority, and
  equivalence designs}.{\BBCQ}
\newblock
\APACjournalVolNumPages{BMC Medical Research Methodology}{19}{}{}.
\newblock
\begin{APACrefDOI} \doi{10.1186/s12874-019-0699-7} \end{APACrefDOI}
\PrintBackRefs{\CurrentBib}

\bibitem [\protect \citeauthoryear {%
{van Schie}%
, {Van Veen}%
, Engelhard%
, Klugkist%
\BCBL {}\ \BBA {} {Van den Hout}%
}{%
{van Schie}%
\ \protect \BOthers {.}}{%
{\protect \APACyear {2016}}%
}]{%
vanSchie:2016}
\APACinsertmetastar {%
vanSchie:2016}%
\begin{APACrefauthors}%
{van Schie}, K.%
, {Van Veen}, S.%
, Engelhard, I.%
, Klugkist, I.%
\BCBL {}\ \BBA {} {Van den Hout}, M.%
\end{APACrefauthors}%
\unskip\
\newblock
\APACrefYearMonthDay{2016}{}{}.
\newblock
{\BBOQ}\APACrefatitle {Blurring Emotional Memories Using Eye Movements:
  Individual Differences and Speed of Eye Movements} {Blurring emotional
  memories using eye movements: Individual differences and speed of eye
  movements}.{\BBCQ}
\newblock
\APACjournalVolNumPages{European Journal of Psychotraumatology}{7}{}{}.
\PrintBackRefs{\CurrentBib}

\bibitem [\protect \citeauthoryear {%
Verdinelli%
\ \BBA {} Wasserman%
}{%
Verdinelli%
\ \BBA {} Wasserman%
}{%
{\protect \APACyear {1995}}%
}]{%
Verdinelli:1995}
\APACinsertmetastar {%
Verdinelli:1995}%
\begin{APACrefauthors}%
Verdinelli, I.%
\BCBT {}\ \BBA {} Wasserman, L.%
\end{APACrefauthors}%
\unskip\
\newblock
\APACrefYearMonthDay{1995}{}{}.
\newblock
{\BBOQ}\APACrefatitle {Computing Bayes Factors Using a Generalization of the
  Savage-Dickey Density Ratio} {Computing bayes factors using a generalization
  of the savage-dickey density ratio}.{\BBCQ}
\newblock
\APACjournalVolNumPages{Journal of American Statistical
  Association}{90}{430}{614-618}.
\newblock
\begin{APACrefDOI} \doi{10.2307/2291073} \end{APACrefDOI}
\PrintBackRefs{\CurrentBib}

\bibitem [\protect \citeauthoryear {%
Vrinten%
\ \protect \BOthers {.}}{%
Vrinten%
\ \protect \BOthers {.}}{%
{\protect \APACyear {2016}}%
}]{%
Vrinten:2016}
\APACinsertmetastar {%
Vrinten:2016}%
\begin{APACrefauthors}%
Vrinten, C.%
, Gu, X.%
, %
, Weinreich, S.%
, Schipper, M.%
, Wessels, J.%
\BDBL {}Verschuuren, J.%
\end{APACrefauthors}%
\unskip\
\newblock
\APACrefYearMonthDay{2016}{}{}.
\newblock
{\BBOQ}\APACrefatitle {An n-of-one RCT for Intravenous Immunoglobulin G for
  Inflammation in Hereditary Neuropathy with Liability to Pressure Palsy
  (HNPP)} {An n-of-one rct for intravenous immunoglobulin g for inflammation in
  hereditary neuropathy with liability to pressure palsy (hnpp)}.{\BBCQ}
\newblock
\APACjournalVolNumPages{Journal of Neurology, Neurosurgery and
  Psychiatry}{87}{7}{790--791}.
\newblock
\begin{APACrefDOI} \doi{10.1136/jnnp-2014-309427} \end{APACrefDOI}
\PrintBackRefs{\CurrentBib}

\bibitem [\protect \citeauthoryear {%
E.~Wagenmakers%
\ \protect \BOthers {.}}{%
E.~Wagenmakers%
\ \protect \BOthers {.}}{%
{\protect \APACyear {2018}}%
}]{%
Wagenmakers:2018}
\APACinsertmetastar {%
Wagenmakers:2018}%
\begin{APACrefauthors}%
Wagenmakers, E.%
, Marsman, M.%
, Jamil, T.%
, Ly, A.%
, Verhagen, J.%
, Love, J.%
\BDBL {}Morey, R.%
\end{APACrefauthors}%
\unskip\
\newblock
\APACrefYearMonthDay{2018}{}{}.
\newblock
{\BBOQ}\APACrefatitle {Bayesian Inference for Psychology. {P}art I:
  {T}heoretical Advantages and Practical Ramifications} {Bayesian inference for
  psychology. {P}art i: {T}heoretical advantages and practical
  ramifications}.{\BBCQ}
\newblock
\APACjournalVolNumPages{Psychonomic Bulletin \& Review}{25}{1}{35--57}.
\newblock
\begin{APACrefDOI} \doi{10.3758/s13423-017-1343-3} \end{APACrefDOI}
\PrintBackRefs{\CurrentBib}

\bibitem [\protect \citeauthoryear {%
E\BHBI J.~Wagenmakers%
}{%
E\BHBI J.~Wagenmakers%
}{%
{\protect \APACyear {2007}}%
}]{%
Wagenmakers:2007}
\APACinsertmetastar {%
Wagenmakers:2007}%
\begin{APACrefauthors}%
Wagenmakers, E\BHBI J.%
\end{APACrefauthors}%
\unskip\
\newblock
\APACrefYearMonthDay{2007}{}{}.
\newblock
{\BBOQ}\APACrefatitle {A Practical Solution to the Pervasive Problem of p
  Values} {A practical solution to the pervasive problem of p values}.{\BBCQ}
\newblock
\APACjournalVolNumPages{Psychonomic Bulletin and Review}{14}{5}{779--804}.
\newblock
\begin{APACrefDOI} \doi{10.3758/BF03194105} \end{APACrefDOI}
\PrintBackRefs{\CurrentBib}

\bibitem [\protect \citeauthoryear {%
E\BHBI J.~Wagenmakers%
, Wetzels%
, Borsboom%
\BCBL {}\ \BBA {} {van der Maas}%
}{%
E\BHBI J.~Wagenmakers%
\ \protect \BOthers {.}}{%
{\protect \APACyear {2017}}%
}]{%
Wagenmakers:2017b}
\APACinsertmetastar {%
Wagenmakers:2017b}%
\begin{APACrefauthors}%
Wagenmakers, E\BHBI J.%
, Wetzels, R.%
, Borsboom, D.%
\BCBL {}\ \BBA {} {van der Maas}, H.%
\end{APACrefauthors}%
\unskip\
\newblock
\APACrefYearMonthDay{2017}{}{}.
\newblock
{\BBOQ}\APACrefatitle {Why Psychologists Must Change the Way They Analyze Their
  Data: {T}he Case of Psi: Comment on Bem (2011)} {Why psychologists must
  change the way they analyze their data: {T}he case of psi: Comment on bem
  (2011)}.{\BBCQ}
\newblock
\APACjournalVolNumPages{Journal of Personality and Social
  Psychology}{100}{}{426--432}.
\newblock
\begin{APACrefDOI} \doi{10.1037/a0022790} \end{APACrefDOI}
\PrintBackRefs{\CurrentBib}

\bibitem [\protect \citeauthoryear {%
Well%
, Kolk%
\BCBL {}\ \BBA {} Klugkist%
}{%
Well%
\ \protect \BOthers {.}}{%
{\protect \APACyear {2008}}%
}]{%
VanWell:2008}
\APACinsertmetastar {%
VanWell:2008}%
\begin{APACrefauthors}%
Well, S\BPBI V.%
, Kolk, A.%
\BCBL {}\ \BBA {} Klugkist, I.%
\end{APACrefauthors}%
\unskip\
\newblock
\APACrefYearMonthDay{2008}{}{}.
\newblock
{\BBOQ}\APACrefatitle {Effects of Sex, Gender Role Identification, and Gender
  relevance of Two Types of Stressors on Cardiovascular and Subjective
  Responses: Sex and Gender Match/Mismatch Effects} {Effects of sex, gender
  role identification, and gender relevance of two types of stressors on
  cardiovascular and subjective responses: Sex and gender match/mismatch
  effects}.{\BBCQ}
\newblock
\APACjournalVolNumPages{Behavior Modification}{32}{}{427--449}.
\PrintBackRefs{\CurrentBib}

\bibitem [\protect \citeauthoryear {%
Westfall%
\ \BBA {} G\"{o}nen%
}{%
Westfall%
\ \BBA {} G\"{o}nen%
}{%
{\protect \APACyear {1996}}%
}]{%
Westfall:1996}
\APACinsertmetastar {%
Westfall:1996}%
\begin{APACrefauthors}%
Westfall, P.%
\BCBT {}\ \BBA {} G\"{o}nen, M.%
\end{APACrefauthors}%
\unskip\
\newblock
\APACrefYearMonthDay{1996}{}{}.
\newblock
{\BBOQ}\APACrefatitle {Asymptotic Properties of ANOVA Bayes Factors}
  {Asymptotic properties of anova bayes factors}.{\BBCQ}
\newblock
\APACjournalVolNumPages{Communications in Statistics: Theory and
  Methods}{25}{}{3101-3123}.
\newblock
\begin{APACrefDOI} \doi{10.1080/03610929608831888} \end{APACrefDOI}
\PrintBackRefs{\CurrentBib}

\bibitem [\protect \citeauthoryear {%
Wetzels%
, Grasman%
\BCBL {}\ \BBA {} Wagenmakers%
}{%
Wetzels%
\ \protect \BOthers {.}}{%
{\protect \APACyear {2010}}%
}]{%
Wetzels:2010}
\APACinsertmetastar {%
Wetzels:2010}%
\begin{APACrefauthors}%
Wetzels, R.%
, Grasman, R\BPBI P\BPBI P\BPBI P.%
\BCBL {}\ \BBA {} Wagenmakers, E\BPBI J.%
\end{APACrefauthors}%
\unskip\
\newblock
\APACrefYearMonthDay{2010}{}{}.
\newblock
{\BBOQ}\APACrefatitle {An Encompassing Prior Generalization of the
  {S}avage-{D}ickey Density Ratio Test} {An encompassing prior generalization
  of the {S}avage-{D}ickey density ratio test}.{\BBCQ}
\newblock
\APACjournalVolNumPages{Computational Statistics and Data
  Analysis}{38}{}{666-690}.
\newblock
\begin{APACrefDOI} \doi{10.1.1.149.885} \end{APACrefDOI}
\PrintBackRefs{\CurrentBib}

\bibitem [\protect \citeauthoryear {%
Wilson%
\ \BBA {} Rule%
}{%
Wilson%
\ \BBA {} Rule%
}{%
{\protect \APACyear {2015}}%
}]{%
Wilson:2015}
\APACinsertmetastar {%
Wilson:2015}%
\begin{APACrefauthors}%
Wilson, J.%
\BCBT {}\ \BBA {} Rule, N.%
\end{APACrefauthors}%
\unskip\
\newblock
\APACrefYearMonthDay{2015}{}{}.
\newblock
{\BBOQ}\APACrefatitle {Facial Trustworthiness Predicts Extreme
  Criminal-Sentencing Outcomes} {Facial trustworthiness predicts extreme
  criminal-sentencing outcomes}.{\BBCQ}
\newblock
\APACjournalVolNumPages{Psychological Science}{26}{}{1325--1331}.
\newblock
\begin{APACrefDOI} \doi{10.1177/0956797615590992} \end{APACrefDOI}
\PrintBackRefs{\CurrentBib}

\bibitem [\protect \citeauthoryear {%
Zondervan-Zwijnenburg%
\ \protect \BOthers {.}}{%
Zondervan-Zwijnenburg%
\ \protect \BOthers {.}}{%
{\protect \APACyear {2019}}%
}]{%
Zondervan:2019}
\APACinsertmetastar {%
Zondervan:2019}%
\begin{APACrefauthors}%
Zondervan-Zwijnenburg, M.%
, Veldkamp, S.%
, Neumann, A.%
, Barzeva, S.%
, Nelemans, S.%
, {van Beijsterveldt}, C.%
\BDBL {}Boomsma, A\BPBI O\BPBI D.%
\end{APACrefauthors}%
\unskip\
\newblock
\APACrefYearMonthDay{2019}{}{}.
\newblock
{\BBOQ}\APACrefatitle {Parental Age and Offspring Childhood Mental Health: A
  Multi-Cohort, Population-Based Investigation} {Parental age and offspring
  childhood mental health: A multi-cohort, population-based
  investigation}.{\BBCQ}
\newblock
\APACjournalVolNumPages{Child Development}{}{}{}.
\newblock
\begin{APACrefDOI} \doi{10.1111/cdev.13267} \end{APACrefDOI}
\PrintBackRefs{\CurrentBib}

\end{thebibliography}

\end{document}